\documentclass[pdflatex,sn-mathphys-num,iicol]{sn-jnl}

\usepackage{graphicx}%
\usepackage{multirow}%
\usepackage{amsmath,amssymb,amsfonts}%
\usepackage{amsthm}%
\usepackage{xcolor}%
\usepackage{textcomp}%
\usepackage{booktabs}%
\usepackage{array}%
\usepackage{algorithm}%
\usepackage{algorithmicx}%
\usepackage{algpseudocode}%
\usepackage{tikz}%

\graphicspath{{figures/}{./}}
\usepackage{pifont}%

\newcommand{\cmark}{\ding{51}}
\newcommand{\xmark}{\ding{55}}
\newcommand{\hd}[1]{{\renewcommand{\arraystretch}{0.95}\begin{tabular}[b]{@{}>{\fontsize{6.8}{7.8}\selectfont}c@{}}#1\end{tabular}}}
\newcommand{\bTheta}{\boldsymbol{\Theta}}
\newcommand{\bx}{\mathbf{x}}
\newcommand{\bh}{\mathbf{h}}
\newcommand{\bW}{\mathbf{W}}

\begin{document}

\title[BrownoutMoE: Structure-Aware Expert Grouping for MoE Serving]{BrownoutMoE: Learning Structure-Aware Expert Grouping for Adaptive Mixture-of-Experts Model Serving}

\author[1,2]{\fnm{Yi} \sur{Ding}}

\author*[1]{\fnm{Minxian} \sur{Xu}}\email{mx.xu@siat.ac.cn}

\author[3]{\fnm{Zhengxin} \sur{Fang}}

\author[1]{\fnm{Kejiang} \sur{Ye}}

\author[4]{\fnm{Chengzhong} \sur{Xu}}

\affil[1]{\orgdiv{Shenzhen Institutes of Advanced Technology}, \orgname{Chinese Academy of Sciences}, \orgaddress{\street{1068 Xueyuan Avenue}, \city{Shenzhen}, \postcode{518055}, \country{China}}}

\affil[2]{\orgname{University of Chinese Academy of Sciences}, \orgaddress{\street{19A Yuquan Road}, \city{Beijing}, \postcode{100049}, \country{China}}}

\affil[3]{\orgname{Victoria University of Wellington}, \orgaddress{\street{PO Box 600}, \city{Wellington}, \postcode{6140}, \country{New Zealand}}}

\affil[4]{\orgdiv{Institute of AI and Brain Sciences, Department of Computer Science}, \orgname{University of Macau}, \orgaddress{\street{Avenida da Universidade, Taipa}, \city{Macau}, \country{China}}}

\abstract{Mixture-of-Experts (MoE) large language models (LLMs) are increasingly served on multi-GPU clusters in cloud data centers, where inference must stay accurate and responsive under bursty demand. Expert access in MoE inference is highly imbalanced: a few hot experts process most tokens, whereas many cold experts fragment kernels and leave GPU parallelism underused. Existing serving systems optimize runtime execution but keep the original expert organization. We propose BrownoutMoE, a structure-aware serving framework inspired by the brownout paradigm in cloud computing. It learns layer-wise expert grouping offline with reinforcement learning, using the post-distillation error as the reward. Grouping-consistent distillation trains one united expert per group, and an online controller adjusts how often united experts replace cold experts under latency objectives. On eight A100 GPUs, BrownoutMoE reduces accuracy degradation by up to 71.4\% relative to index-order grouping, improves unfused throughput by up to 2.24$\times$ over vLLM, and lowers saturated decode latency by 38\%.}


\makeatletter
\def\corrauthemail{\textcolor{blue}{mx.xu@siat.ac.cn}}
\makeatother

\maketitle

\section{Introduction}\label{sec:intro}

Large language models (LLMs) are now embedded in online services such as search, recommendation, question answering, fact checking~\cite{mandon2026halifacts}, and interactive agents~\cite{yu2022orca}. These services are hosted on servers and clusters equipped with multiple graphics processing units (GPUs) in cloud data centers, which places inference on an online request path rather than an offline batch job. The serving system must absorb bursty arrivals, heterogeneous prompt lengths, and streaming outputs. It must also control tail latency and preserve answer quality across long-lived sessions~\cite{dean2013tail}, while keeping expensive accelerators highly utilized.

Mixture-of-Experts (MoE) models~\cite{shazeer2017sparsely,jiang2024mixtral,liu2024deepseek} are attractive because they provide large parameter capacity while activating only a small subset of feed-forward experts for each token. Sparse activation, however, does not imply balanced execution. Routing can produce a long-tailed access distribution in which a small set of hot experts receives most tokens~\cite{zhou2022expertchoice,zoph2022stmoe}. When experts are distributed across GPUs, hot experts become synchronization points under bursty request traffic, whereas cold experts create fragmented kernels and underutilize memory and interconnect bandwidth.

Most existing MoE serving work addresses this problem at execution time. Expert parallelism, adaptive runtimes, and block-sparse or fused kernels improve communication and arithmetic efficiency~\cite{rajbhandari2022deepspeedmoe,hwang2023tutel,gale2023megablocks}. Request scheduling and prefill/decode disaggregation target queueing and phase imbalance~\cite{yu2022orca,zhong2024distserve,agrawal2024sarathi,patel2024splitwise}. These techniques generally retain the original expert inventory and layout. They therefore do not remove the structural cost of repeatedly loading and executing rarely used experts. BrownoutServe~\cite{11357542} takes a complementary approach by grouping original experts into a distilled \emph{united expert} (a single student network that approximates the functions of its member experts) and selecting between original and united experts with a brownout threshold driven by service-level objectives (SLOs). Its remaining limitation is that index-order grouping is independent of functional similarity between experts.

\begin{figure*}[t]
\centering
\includegraphics[width=0.98\textwidth]{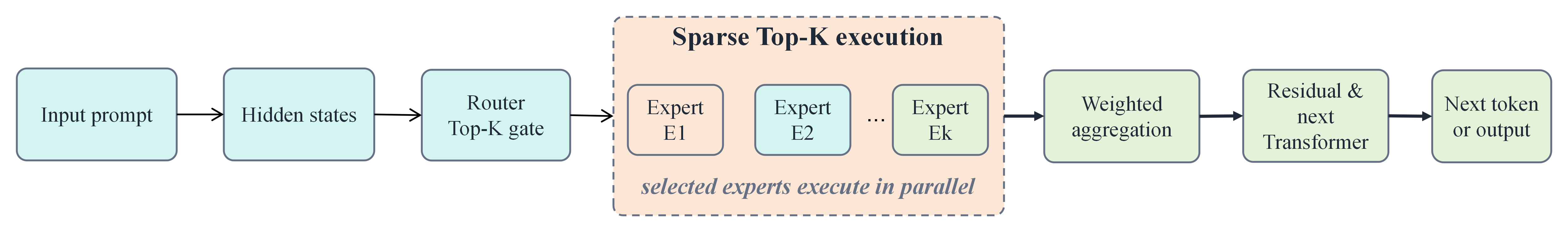}
\caption{MoE inference flow. The router selects a sparse Top-$K$ subset, the selected experts execute in parallel inside the sparse execution stage, and their outputs are aggregated before the next Transformer block}
\label{fig:moe-inference}
\end{figure*}

Figure~\ref{fig:moe-inference} summarizes the inference path that motivates our design. For each token, the router selects only a small Top-$K$ subset of experts, the selected experts execute in parallel, and their outputs are combined before the next Transformer block. Although sparse activation reduces arithmetic for an individual token, dispatching tokens to many independently used experts and executing each of them separately can still create fragmented kernels and communication hot spots. BrownoutMoE addresses this structural cost by learning which experts can share a united expert while leaving the router and the online request interface unchanged.

The grouping decision is consequential because distillation must approximate a mixture of expert functions with a single student. Grouping behaviorally similar experts gives the student a coherent target. Grouping dissimilar experts forces one parameter set to cover incompatible activation regions. This issue is related to recent work on expert pruning, merging, and compression~\cite{lu2024naee,liu2024eep,guo2025cprune,li2025submoe}, which also reports quality loss when specialized functions are combined without a behavior-aware objective. The problem becomes more severe for aggressive grouping factors, because approximation error can compound across layers. A grouping algorithm should therefore optimize post-distillation behavior rather than a proxy based only on expert indices or parameter distance.

This paper proposes BrownoutMoE, a structure-aware extension of the brownout serving paradigm. For each MoE layer, it learns a capacity-constrained map from original experts to united groups. A calibration pass records routing frequencies and expert outputs. Hierarchical clustering provides a warm start, after which Group Relative Policy Optimization (GRPO)~\cite{shao2024deepseekmath,schulman2017ppo} explores local group swaps. We adapt the GRPO principle to constrained combinatorial expert grouping. Each candidate is evaluated by short-horizon united-expert distillation. Its routing-weighted post-distillation mean squared error (MSE) becomes the reward. The selected map is then used in a full distillation pass and exported with the united-expert weights.

BrownoutMoE separates the offline structure problem from the online control problem. The offline module searches for a reusable grouping artifact and trains the corresponding students. At request time, the original router is unchanged. The runtime only consults the saved map for eligible cold experts. The SLO Analyzer independently adjusts the brownout threshold from time-to-first-token (TTFT), time-per-output-token (TPOT), and tail latency. This separation is compatible with continuous batching, FlashAttention~\cite{dao2022flashattention,dao2023flashattention2}, PagedAttention~\cite{kwon2023efficient}, and fused MoE operators. GRPO therefore adds no request-path computation.

Our study addresses three practical questions: (i) whether behavior-aware grouping reduces the fidelity loss of united experts, (ii) whether the gain remains when the grouping factor increases, and (iii) whether the grouping artifact improves throughput without weakening SLO control. On Qwen1.5-MoE-A2.7B served with tensor parallelism on eight A100 GPUs, BrownoutMoE reduces accuracy degradation by up to 71.4\% relative to sequential grouping, and its brownout path improves unfused throughput by up to 2.24$\times$ over a vLLM baseline, while preserving the same runtime interface.

The key contributions of this paper are as follows:
\begin{itemize}
    \item We formulate layer-wise expert grouping as a discrete policy optimization problem and use the true routing-weighted post-distillation MSE as the GRPO reward.
    \item We develop a two-phase optimization pipeline that combines similarity-based warm-start, swap-based GRPO search, capacity constraints, and grouping-consistent united-expert distillation.
    \item We integrate the learned grouping artifact with an SLO-aware brownout serving path and evaluate accuracy, throughput, multi-GPU scalability, and latency on a tensor-parallel deployment with up to eight A100 GPUs under LLM service workloads.
\end{itemize}

The remainder of this paper is organized as follows. Section~\ref{sec:background} introduces the operating principles of MoE inference services, the united-expert brownout mechanism, and the load and accuracy challenges that motivate structure-aware grouping. Section~\ref{sec:related} reviews related work on MoE execution systems, routing and load balancing, expert-structure compression, runtime expert management, and adaptive quality control, and positions BrownoutMoE against these directions. Section~\ref{sec:design} presents the system architecture, problem formulation, GRPO-based grouping search, grouping-consistent distillation, and the SLO-aware online controller, including their algorithms and complexity. Section~\ref{sec:eval} reports performance evaluations on downstream accuracy, throughput, scalability, and latency, while Section~\ref{sec:discussion} discusses practical implications and limitations. Section~\ref{sec:conclusion} concludes the paper and outlines future work.

\section{Background and Motivation}\label{sec:background}

In this section, we briefly introduce MoE-based LLMs and the united expert mechanism. We then analyze the load imbalance, accuracy degradation, and burst-handling problems that motivate our approach.

\subsection{Mixture-of-Experts Inference Services}

MoE-based LLMs~\cite{shazeer2017sparsely,fedus2022switch,jiang2024mixtral} replace the feed-forward networks (FFNs) in standard Transformers with a gating function and multiple small FFNs (experts). Each incoming token is routed to a subset of experts, reducing per-token computation and active parameters. For an MoE layer with $E$ experts and top-$K$ routing, the output for token $\bx_t$ is $\bh_t = \sum_{i=1}^{K} g_i(\bx_t) \cdot \mathrm{FFN}_i(\bx_t)$, where $g_i$ is the gating score. MoE inference has two stages: prefill processes the prompt in parallel, whereas decoding generates output tokens one at a time. In multi-GPU deployments, expert weights are partitioned across devices through tensor or expert parallelism, so the routed tokens of each layer are dispatched to, and gathered from, the GPUs that hold the selected experts. Practical deployment therefore remains sensitive to expert load imbalance and communication overhead.

\subsection{United Expert Model and Brownout Approach}\label{sec:united}

To address the MoE inference bottleneck, BrownoutServe~\cite{11357542} introduced \emph{united experts} and the \emph{brownout approach}, inspired by brownout control for adaptive cloud applications~\cite{klein2014brownout,xu2021selfadaptive}. The key idea is to group multiple original experts into one united expert through knowledge distillation~\cite{hinton2015distilling}, reducing the number of experts loaded during inference. The approach provides three operating modes: \emph{zero-brownout} keeps all experts active; \emph{full-brownout} routes all tokens through united experts; and \emph{partial-brownout} delegates only cold-expert tokens to united experts using a configurable threshold.

\textbf{Features and limitations.} While the united expert mechanism reduces memory and communication overhead with flexible latency control, BrownoutServe assigns experts to groups by index order, ignoring behavioral relationships. Figure~\ref{fig:motivation}(a) illustrates the consequence with two MoE layers, each containing eight original experts (E0--E7) that are partitioned into two united experts (UE1 and UE2) of four members; experts shown in the same color produce similar outputs. BrownoutServe groups experts sequentially, so UE1 always contains E0--E3 and UE2 contains E4--E7. In Layer1, the dissimilar experts E2 and E3 are therefore forced into the same united expert, while the similar pairs $\{$E2, E5$\}$ and $\{$E3, E4$\}$ are split across UE1 and UE2. BrownoutMoE instead regroups experts by behavioral similarity: in Layer1 it places E3 with E4 in UE1 and E2 with E5 in UE2, and in Layer2 it forms a different map that keeps $\{$E1, E6$\}$ in UE1 and $\{$E0, E7$\}$ in UE2. Each united expert therefore approximates a more homogeneous set of functions, and the map can differ from layer to layer. This mismatch in sequential grouping causes accuracy degradation that worsens as the grouping factor $k$ (the number of original experts per united expert) increases, motivating a structure-aware grouping strategy.

\subsection{Challenges in Structure-Aware Brownout Serving}\label{sec:challenges}

Although the brownout approach reduces inference latency, three challenges remain: expert load imbalance, accuracy degradation caused by suboptimal expert grouping, and static brownout control under bursty demand.

\textbf{C1: Expert load imbalance.}
Token distribution across experts exhibits a pronounced long-tail pattern: a small portion of experts handles most tokens, whereas the majority remains underutilized. This imbalance is a recognized challenge in sparse-model training~\cite{zhou2022expertchoice,zoph2022stmoe}. Serving systems additionally report communication and dispatch bottlenecks caused by uneven expert demand~\cite{hwang2023tutel,he2026banaserve}. On Qwen1.5-MoE-A2.7B, the top-10 experts handle over 40\% of routed tokens across 24 MoE layers, while the bottom-20 collectively process less than 10\%. Grouping quality therefore affects hot experts disproportionately. Placing frequently activated experts in one group raises the united expert's traffic, whereas grouping cold experts wastes distillation capacity on rarely observed combinations.

\textbf{C2: Accuracy loss from sequential grouping.}
As discussed in Section~\ref{sec:united}, BrownoutServe~\cite{11357542} uses sequential grouping and ignores expert behavioral similarity. We empirically validate this effect by computing the $60 \times 60$ pairwise cosine-similarity matrix of expert outputs on Qwen1.5-MoE-A2.7B and applying hierarchical clustering. Clustering-based grouping reduces distillation MSE by 17.9\%--48.8\% across MoE layers compared with sequential grouping. This result confirms that grouping quality directly affects united-expert fidelity.

\begin{figure*}[t]
\centering
\begin{minipage}[t]{0.58\textwidth}
\centering
\includegraphics[width=\linewidth]{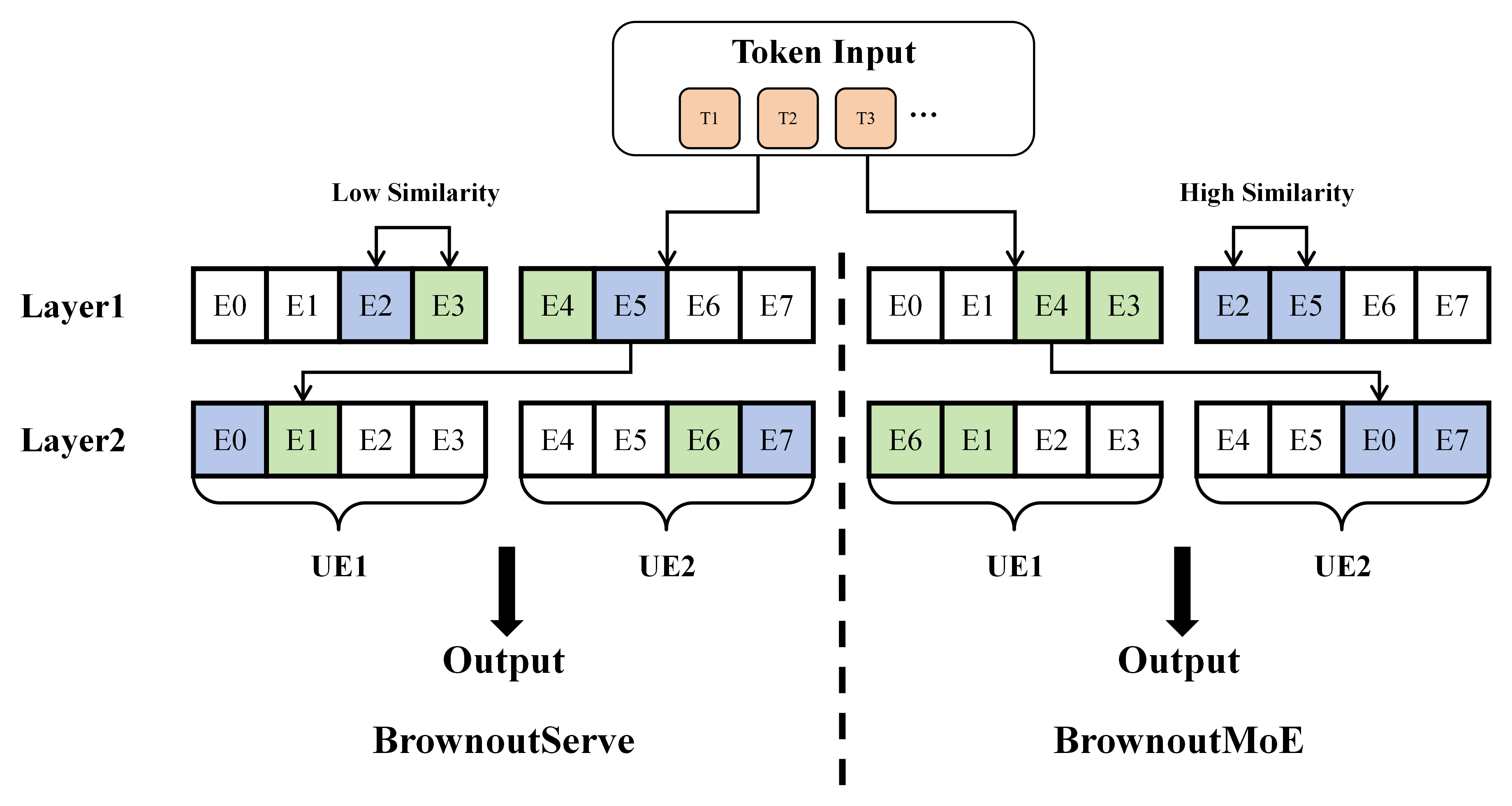}\\
\small (a)
\end{minipage}
\hfill
\begin{minipage}[t]{0.38\textwidth}
\centering
\includegraphics[width=\linewidth]{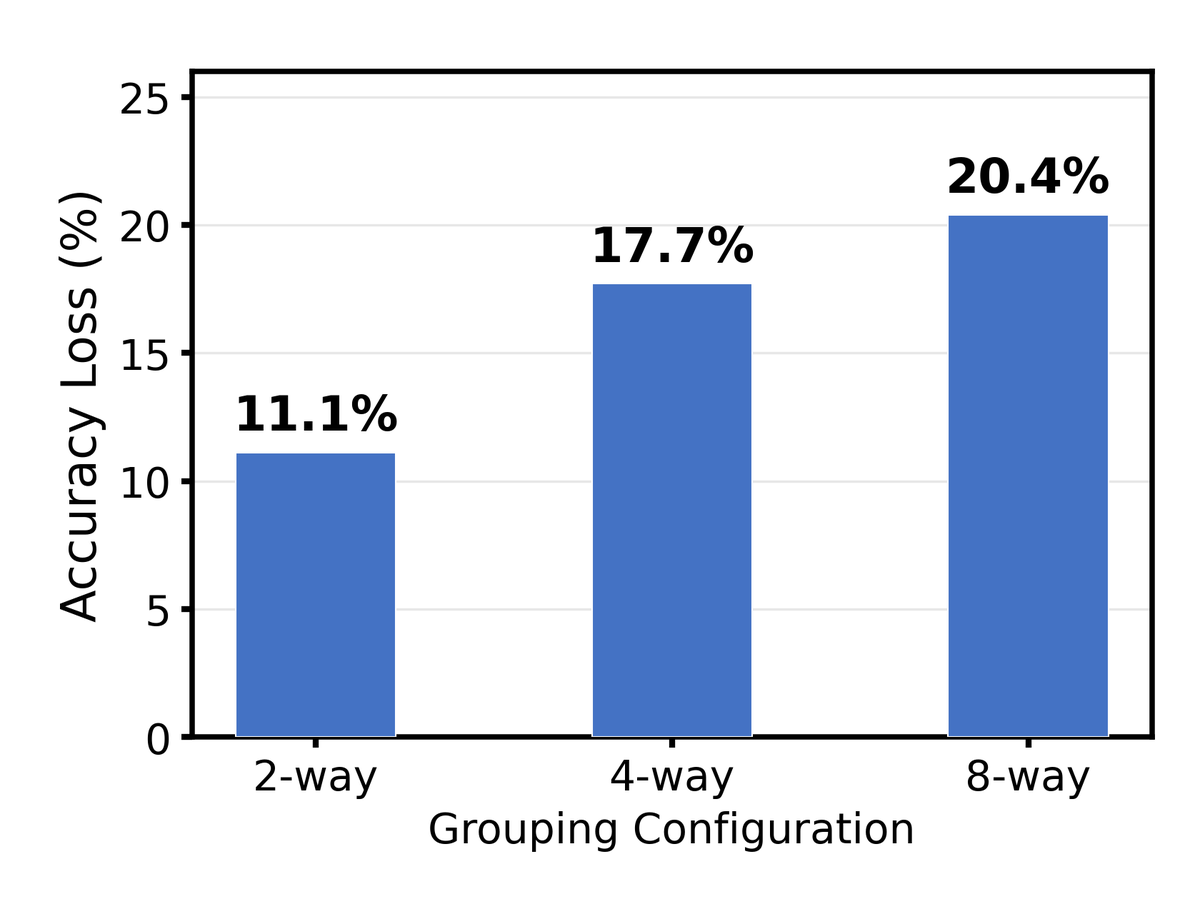}\\
\small (b)
\end{minipage}
\caption{Motivation for structure-aware grouping. (a) Expert grouping in BrownoutServe and BrownoutMoE for two MoE layers with eight experts each; experts in the same color have similar outputs, and UE denotes a united expert. (b) C-Eval accuracy loss under sequential grouping with different grouping configurations}
\label{fig:motivation}
\end{figure*}

Moreover, the gap widens as the grouping factor $k$ increases: under 8-way grouping, the dissimilarity within sequentially formed groups becomes more severe, leading to larger accuracy drops on downstream tasks. We further quantify this on the C-Eval benchmark: the accuracy loss from zero brownout to full brownout grows from 11.1\% under 2-way grouping to 20.4\% under 8-way grouping, as shown in Figure~\ref{fig:motivation}(b). This monotonic trend confirms that as more experts are merged into each group, the grouping strategy becomes increasingly critical for preserving model accuracy.

\begin{figure*}[t]
\centering
\includegraphics[width=0.95\textwidth]{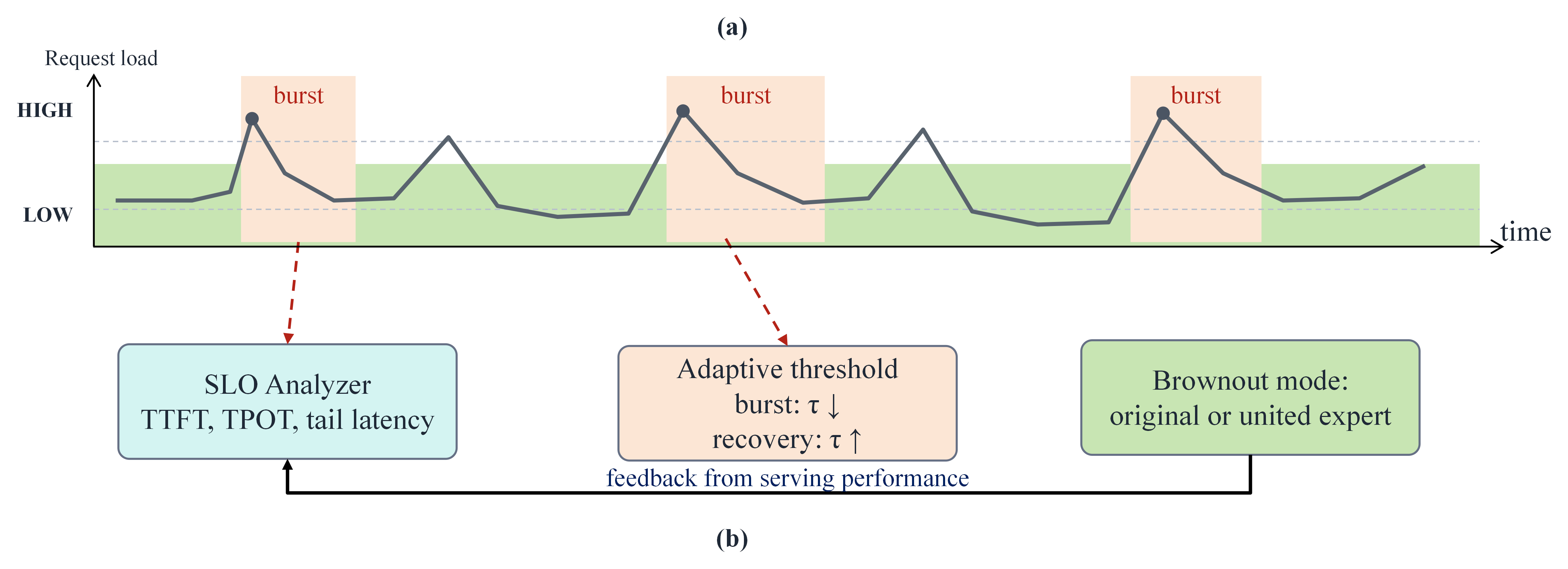}
\caption{Dynamic brownout control for bursty MoE serving. (a) Request load over time, where shaded burst intervals push the load above the high level. (b) Online SLO control loop: the SLO Analyzer observes TTFT, TPOT, and tail latency, the adaptive threshold lowers $\tau$ during a burst and raises it during recovery, and the brownout mode selects original or united experts, with serving performance fed back to the analyzer}
\label{fig:motivation-control}
\end{figure*}

\textbf{C3: Static brownout cannot track bursty demand.}
Even a well-formed grouping map cannot determine how aggressively to use united experts for every workload phase due to the dynamic features of workloads~\cite{Wen2024}. A fixed threshold either leaves performance on the table during a burst or sacrifices accuracy after the workload has returned to normal. The service therefore needs a lightweight online controller that reacts to TTFT, TPOT, and tail latency while preserving the offline grouping artifact. Figure~\ref{fig:motivation-control} makes this control path explicit. Figure~\ref{fig:motivation-control}(a) shows a request load that stays in a normal band most of the time but is interrupted by short burst intervals that push it above the high level. Figure~\ref{fig:motivation-control}(b) shows the online control loop that responds to these bursts: the SLO Analyzer observes TTFT, TPOT, and tail latency; the adaptive controller lowers the brownout threshold $\tau \in [0,1]$---the fraction of eligible cold-expert calls kept on the original experts---during a burst so that more united experts are used, and raises it again during recovery; and the brownout mode applies the resulting choice between original and united experts, whose serving performance is fed back to the analyzer.

\subsection{Motivation}

The analysis of C1 and C2 reveals two key insights. First, the quality of expert grouping has a direct and measurable impact on distillation accuracy: grouping behaviorally similar experts yields significantly lower MSE than naive sequential grouping, as confirmed by the 17.9\% to 48.8\% MSE reduction observed through hierarchical clustering on Qwen1.5-MoE-A2.7B. Second, the grouping problem is inherently combinatorial: with 60 experts partitioned into $G$ groups, the search space grows factorially, making exhaustive enumeration infeasible even for moderate grouping factors.

These observations motivate a learning-based approach. Rather than relying on fixed heuristics, we formulate expert grouping as a discrete policy optimization problem where a learned policy assigns experts to groups. The grouping quality is measured by the true post-distillation MSE after short-horizon united-expert distillation, which serves as the reward signal and directly reflects the deployment objective of preserving original model behavior.

We adopt Group Relative Policy Optimization (GRPO)~\cite{shao2024deepseekmath,schulman2017ppo} for this task. GRPO generates candidate groupings by perturbing a base grouping through expert-pair swaps. It computes advantages within each perturbation group, reducing variance relative to batch-level normalization. We initialize the policy with hierarchical clustering on expert-output similarities. BrownoutMoE therefore uses a two-phase pipeline: GRPO-based grouping search followed by grouping-consistent united-expert distillation.

The third challenge adds a temporal dimension to the design. Offline grouping determines \emph{which} experts can share a student, while the SLO controller determines \emph{when} that substitution should be used. Separating these decisions allows the expensive structure search to remain offline and keeps the online control loop to a single scalar threshold update.

\section{Related Work}\label{sec:related}

We organize prior work by the layer at which it changes an MoE service. Execution systems reduce the cost of dispatch and communication, routing methods shape the token-to-expert workload, structural methods change the expert inventory, runtime managers decide which weights are resident, and adaptive controllers trade optional computation for quality. This organization separates optimizing \emph{how an expert is executed} from optimizing \emph{which experts can safely share} a student.

\begin{table*}[t]
\caption{Comparison of MoE structure and runtime-management methods; BrownoutMoE combines all listed capabilities}
\label{tab:related-comparison}
\centering
\fontsize{8}{9.5}\selectfont
\renewcommand{\arraystretch}{1.15}
\setlength{\tabcolsep}{0.8pt}
\begin{tabular}{@{}p{73.5pt}>{\centering\arraybackslash}p{32.5pt}>{\centering\arraybackslash}p{25.5pt}>{\centering\arraybackslash}p{33pt}>{\centering\arraybackslash}p{27pt}>{\centering\arraybackslash}p{24pt}>{\centering\arraybackslash}p{33.5pt}>{\centering\arraybackslash}p{29.5pt}>{\centering\arraybackslash}p{25.5pt}>{\centering\arraybackslash}p{26pt}>{\centering\arraybackslash}p{34.5pt}>{\centering\arraybackslash}p{34pt}>{\centering\arraybackslash}p{34.5pt}@{}}
\toprule
 & \multicolumn{7}{c}{\fontsize{7.5}{9}\selectfont\textbf{Offline model/structure}} & \multicolumn{2}{c}{\fontsize{7.5}{9}\selectfont\begin{tabular}[b]{@{}c@{}}\textbf{Online}\\\textbf{serving}\end{tabular}} & \multicolumn{3}{c}{\fontsize{7.5}{9}\selectfont\textbf{Reported impact}} \\
\cmidrule(lr){2-8}\cmidrule(lr){9-10}\cmidrule(lr){11-13}
{\fontsize{7}{8}\selectfont\textbf{Method}} & {\hd{Expert\\grouping}} & {\hd{United\\expert}} & {\hd{Behavior\\aware}} & {\hd{Layer-\\specific\\map}} & {\hd{RL-\\based\\search}} & {\hd{Post-\\distill.\\objective}} & {\hd{Distill./\\retrain}} & {\hd{Online\\mode\\adjust.}} & {\hd{SLO\\control}} & {\hd{Through-\\put/\\latency}} & {\hd{Accuracy}} & {\hd{Memory/\\comm.}} \\
\midrule
NAEE~\cite{lu2024naee} & \xmark & \xmark & \xmark & \cmark & \xmark & \xmark & \xmark & \xmark & \xmark & \cmark & \cmark & \cmark \\
EEP~\cite{liu2024eep} & \cmark & \cmark & \xmark & \cmark & \xmark & \xmark & \xmark & \xmark & \xmark & \cmark & \cmark & \cmark \\
C-Prune~\cite{guo2025cprune} & \cmark & \xmark & \cmark & \cmark & \xmark & \xmark & \xmark & \xmark & \xmark & \cmark & \cmark & \cmark \\
Sub-MoE~\cite{li2025submoe} & \cmark & \cmark & \cmark & \cmark & \xmark & \xmark & \xmark & \xmark & \xmark & \cmark & \cmark & \cmark \\
Huang et al.~\cite{huang2024toward} & \xmark & \xmark & \xmark & \xmark & \xmark & \xmark & \xmark & \cmark & \xmark & \cmark & \xmark & \cmark \\
MoE-Infinity~\cite{xue2024moeinfinity} & \xmark & \xmark & \xmark & \xmark & \xmark & \xmark & \xmark & \cmark & \xmark & \cmark & \xmark & \cmark \\
ExpertFlow~\cite{he2024expertflow} & \xmark & \xmark & \xmark & \xmark & \xmark & \xmark & \xmark & \cmark & \xmark & \cmark & \xmark & \cmark \\
BrownoutServe~\cite{11357542} & \cmark & \cmark & \xmark & \xmark & \xmark & \xmark & \cmark & \cmark & \cmark & \cmark & \cmark & \cmark \\
\midrule
\textbf{BrownoutMoE} & \cmark & \cmark & \cmark & \cmark & \cmark & \cmark & \cmark & \cmark & \cmark & \cmark & \cmark & \cmark \\
\bottomrule
\end{tabular}
\end{table*}

\textbf{MoE execution systems.} Expert parallelism distributes experts across devices and uses collective communication to exchange routed tokens. DeepSpeed-MoE~\cite{rajbhandari2022deepspeedmoe} and Tutel~\cite{hwang2023tutel} reduce dispatch overhead through optimized all-to-all paths, placement strategies, and fused kernels. MegaBlocks~\cite{gale2023megablocks} converts irregular token-expert assignments into block-sparse matrix operations, improving arithmetic utilization without changing the model's expert functions. These systems are orthogonal to BrownoutMoE: they accelerate the execution of the selected experts, but still keep and execute every original expert and therefore do not address the approximation cost of replacing several functions with one united expert.

\textbf{Routing, load balancing, and serving scheduling.} A separate line of work changes the distribution of tokens before execution. Switch Transformer~\cite{fedus2022switch} uses sparse top-one routing and capacity constraints to simplify dispatch, while Expert Choice routing~\cite{zhou2022expertchoice} assigns tokens from the expert side to improve utilization. ST-MoE~\cite{zoph2022stmoe} adds router regularization and stability mechanisms to reduce training-time instability. At the serving layer, Orca~\cite{yu2022orca}, Sarathi-Serve~\cite{agrawal2024sarathi}, and Splitwise~\cite{patel2024splitwise} schedule requests or separate prefill and decode phases to control queueing and device utilization. These methods can reduce load skew or hide communication, but the router still selects from the same expert set; they do not learn a behaviorally compatible united expert for brownout substitution.

\textbf{Expert-structure compression.} Post-training compression methods modify the expert inventory itself. NAEE~\cite{lu2024naee} combines expert pruning with activation skipping, reducing the number of parameters or executions but retaining a pruning-oriented objective. EEP~\cite{liu2024eep} searches over pruning, expert merging, and router transformations, providing a broader structural search but still producing a fixed compressed model. C-Prune~\cite{guo2025cprune} clusters experts using parameter similarity and removes redundant clusters, whereas Sub-MoE~\cite{li2025submoe} uses output similarity and subspace merging to preserve more of the original function. These methods establish that expert structure is a useful compression variable. Their usual deployment point, however, is an offline permanently compressed checkpoint; they do not retain an SLO-aware controller that can vary the substitution level at request time.

\textbf{Runtime expert management.} Runtime systems leave the model structure intact and optimize which experts are loaded or executed under a memory and bandwidth budget. Huang et al.~\cite{huang2024toward} combine dynamic gating, expert buffering, and load balancing so that frequently requested experts remain available on the accelerator. MoE-Infinity~\cite{xue2024moeinfinity} uses request-level routing traces for expert prefetching and cache management, while ExpertFlow~\cite{he2024expertflow} predicts routing paths and schedules tokens to reduce cache misses and transfers. These approaches are effective when the bottleneck is weight movement or cache residency. They are orthogonal to BrownoutMoE because they preserve every original expert function and do not train a student that can replace a group of experts.

\textbf{Adaptive quality control.} Brownout control was introduced for cloud applications in which optional features can be disabled temporarily to keep a service within its latency objective~\cite{klein2014brownout}. Tail-latency studies show that queueing and straggler effects make the high percentile more informative than mean latency for interactive services~\cite{dean2013tail}. BrownoutServe~\cite{11357542} transfers this idea to MoE inference: it distills a united expert, keeps the original router, and dynamically sends eligible cold-expert calls to the united path. Its control loop provides a useful runtime interface, but its index-order grouping does not use expert behavior and can force dissimilar functions into the same student. BrownoutMoE preserves the brownout interface while replacing the fixed grouping with routing-aware, post-distillation optimization, thereby connecting offline structural quality with online quality--latency control.

Table~\ref{tab:related-comparison} separates offline structure changes from online serving control and uses only criteria that are meaningful for MoE inference optimization. A check mark (\cmark) indicates that a method explicitly provides or evaluates the corresponding capability; a cross (\xmark) means that the method does not provide the capability or that it is outside the method's scope. Three offline criteria separate learned grouping from fixed or heuristic grouping: whether the structure decision is made separately for each MoE layer (layer-specific map), whether candidate structures are explored by reinforcement learning (RL), denoted RL-based search, and whether candidates are scored by the error of a distilled student rather than by expert indices or parameter distance (post-distillation objective). BrownoutServe shares the brownout serving path with BrownoutMoE, but it applies the same index-order map to every layer without search or quality feedback; BrownoutMoE therefore differs from it in all four behavior- and learning-related criteria. The final three columns describe reported evaluation targets rather than directly comparable numerical values, because the cited works use different models, hardware, and workloads. Because these categories can coexist in a production stack, the table compares methods by intervention point and direct optimization target rather than by throughput values measured on different hardware. To the best of our knowledge, BrownoutMoE is among the first approaches to combine learned, layer-specific expert grouping with united-expert distillation, dynamic runtime adjustment, and SLO control.

\section{System Design and Methodology}\label{sec:design}

\subsection{System Overview}\label{sec:overview}

\begin{figure*}[t]
\centering
\includegraphics[width=0.95\textwidth]{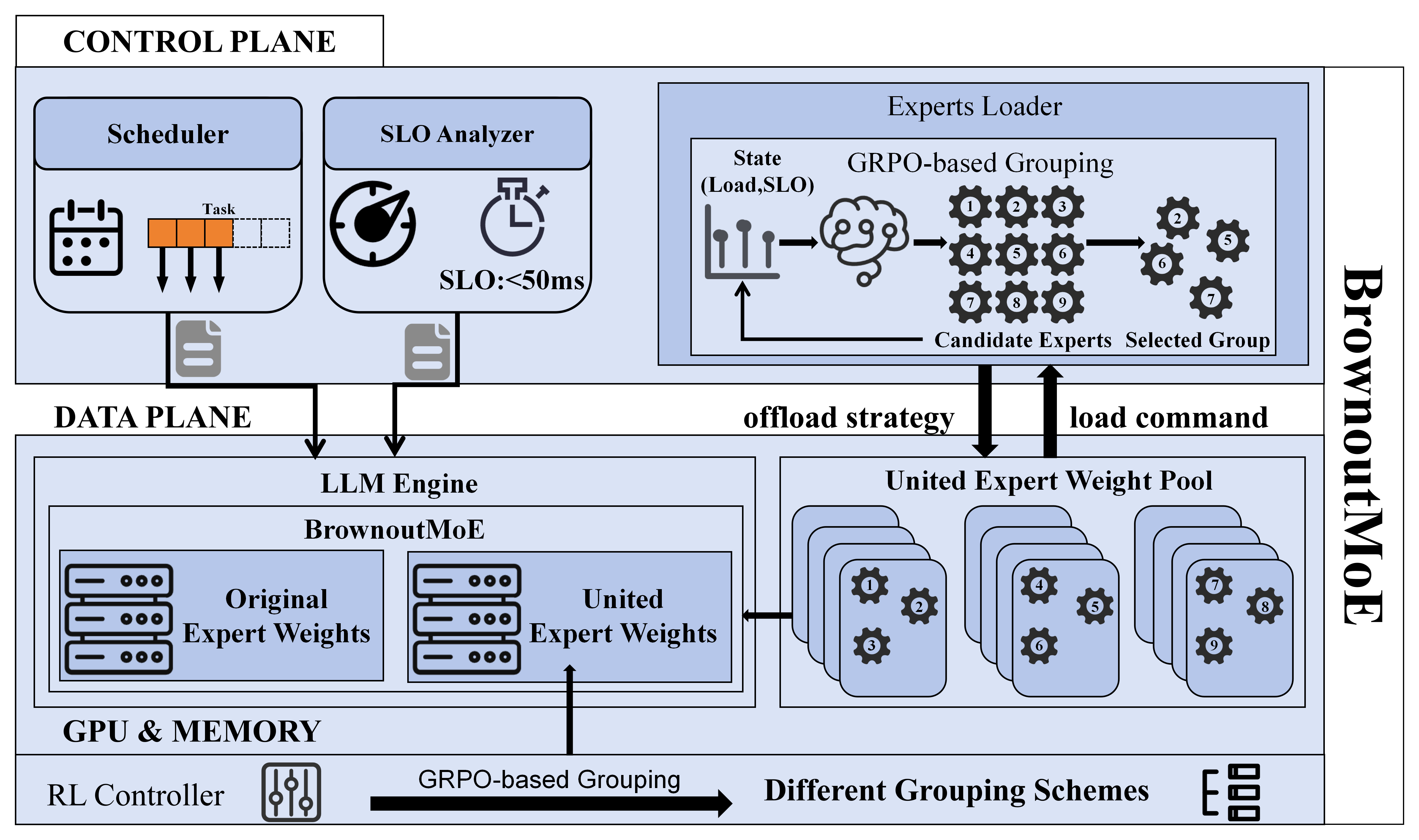}
\caption{Overview of the BrownoutMoE system design, including the control plane, data plane, GPU memory pools, and offline structure-aware grouping path}
\label{fig:overview}
\end{figure*}

BrownoutMoE is an end-to-end structure-aware MoE inference serving framework for efficient and accurate LLM services on multi-GPU servers and clusters. Figure~\ref{fig:overview} shows the complete request and model-management path. Users submit streaming requests to the service front end. The control plane observes the request and model state, while the data plane executes tokens on the GPUs. An offline structure optimizer supplies the grouping artifact consumed by the online components; this division prevents exploratory reinforcement learning from entering the latency-critical request path.

The framework has four cooperating paths: request scheduling, SLO feedback, model-weight management, and structure optimization. The first three execute continuously; the fourth is run during model preparation or when a deployment policy requests a new grouping artifact.

\textbf{Request ingress and scheduling.} The Scheduler receives incoming requests and places them in a continuous-batching queue. First-come, first-served (FCFS) ordering provides predictable admission, while streaming allows a request to release each generated token as soon as it is available. The scheduler records request identifiers, arrival time, prompt length, generation budget, and completion state. These fields are shared with the SLO Analyzer so that queueing delay can be distinguished from model execution time.

\textbf{SLO Analyzer and brownout feedback.} The SLO Analyzer aggregates TTFT, TPOT, 90th- and 99th-percentile (P90/P99) latency, queue length, and GPU utilization over a sliding window. It compares these measurements with the service objective and applies Algorithm~\ref{alg:slo_control} to update one scalar threshold. If tail latency approaches or exceeds the SLO, the threshold is reduced so that more eligible cold-expert calls use united experts. If latency is comfortably below the warning line, the threshold is increased to recover accuracy. The analyzer never changes router weights or the grouping map while a request is executing, which avoids inconsistent routing within a sequence.

\textbf{Experts Loader and GPU memory.} The Experts Loader is responsible for the two weight pools shown in Figure~\ref{fig:overview}: the original expert pool and the united-expert pool. It loads the selected model version, pins the active tensors in GPU memory, and maintains a map from each original expert identifier to its united group. Under memory pressure, the loader can offload inactive original experts and issue a load command for the united weights; the reverse operation restores the zero-brownout path. Weight versions and grouping hashes are checked together, so a map produced for one model checkpoint cannot silently be applied to another.

\textbf{Structure-Aware Grouping Module.} The module is the offline component at the top of the control plane. It runs a calibration dataset through the unchanged gate, records per-layer routing frequencies, and stores compact expert-output summaries. From these summaries it computes pairwise cosine similarities and creates a hierarchical-clustering warm-start. GRPO then samples capacity-constrained groupings, evaluates them with short-horizon distillation, and returns the map with the best routing-weighted MSE. A full distillation pass produces the united-expert weights. The module exports both artifacts as an immutable model-version package consumed by the Experts Loader.

\textbf{LLM engine and BrownoutMoE operator.} The data plane contains the LLM engine, the original and united expert weights, and the GPU memory manager. The engine uses PyTorch with Triton implementations of FlashAttention~\cite{dao2022flashattention,dao2023flashattention2} and PagedAttention~\cite{kwon2023efficient}, together with continuous batching. BrownoutMoE is a pluggable MoE operator. It executes the original gate, looks up the grouping map, and checks the current brownout threshold. Hot experts remain on the original path. For an eligible cold expert, the operator redirects the token to its united expert and combines the result using the original gate weight. The hidden-state interface and residual connection remain unchanged, allowing coexistence with standard Transformer layers.

\textbf{Online and offline interfaces.} The arrows labeled \emph{load command} and \emph{offload strategy} in Figure~\ref{fig:overview} represent a controlled boundary between model management and execution. The online path reads a read-only map and threshold, whereas the offline path writes a new map only after calibration, search, and distillation finish. This boundary supports rollback: the loader can restore the previous package without retraining or restarting the scheduler. It also makes the cost of GRPO measurable as an offline preparation cost rather than a serving overhead.

\textbf{Execution invariants.} BrownoutMoE preserves (i) the original router and top-$K$ probabilities, (ii) the tensor-parallel placement of each layer, (iii) the hidden-state dimensionality of every expert, and (iv) deterministic map lookup for a model version. These invariants ensure that the learned grouping changes only the optional expert computation. They also make comparisons with BrownoutServe meaningful because both systems use the same request scheduler, SLO policy, calibration data, and inference kernels; the controlled variable is the grouping structure.

\subsection{Problem Formulation}\label{sec:formulation}

For each MoE layer containing $E$ routed experts ($E = 60$ for Qwen1.5-MoE-A2.7B), under a $k$-way grouping scheme, the number of groups is:
\begin{equation}
\label{eq:num_groups}
G = \left\lceil \frac{E}{k} \right\rceil.
\end{equation}
When $E$ is not divisible by $k$, the remaining experts form a residual group with fewer than $k$ members. For example, with $E = 60$ experts and $k = 8$, the layer forms $G = \lceil 60/8 \rceil = 8$ united experts: seven full groups of 8 experts and one residual group of 4. A grouping is a mapping:
\begin{equation}
\label{eq:grouping_map}
g : \mathcal{E} \rightarrow \{1, 2, \dots, G\},
\end{equation}
where $G_m = \{e \in \mathcal{E} : g(e) = m\}$ denotes the set of experts assigned to group $m$ by grouping $g$. The goal is to find an optimal grouping $g^*$ that minimizes the per-layer approximation error on representative calibration inputs:
\begin{equation}
\label{eq:objective}
\begin{aligned}
g^* &= \arg\min_{g} \; \mathrm{MSE}_{\text{total}}(g)\\
    &= \arg\min_{g} \sum_{m=1}^{G} \beta_m \cdot \mathrm{MSE}_m(g),
\end{aligned}
\end{equation}
where $\mathrm{MSE}_m(g)$ is the distillation error of the $m$-th united expert trained under grouping $g$, and $\beta_m = \sum_{e \in G_m} w_e$ aggregates the routing frequencies of its member experts. Formally, let $\mathcal{D}$ denote the calibration dataset and let $N_e$ count how often expert $e$ appears in the Top-$K$ routing decisions over all tokens $\bx \in \mathcal{D}$:
\begin{equation}
\label{eq:we}
\begin{gathered}
w_e = \frac{N_e}{\sum_{e' \in \mathcal{E}} N_{e'}},\\
N_e = \sum_{\bx \in \mathcal{D}} \mathbf{1}\!\left[\, e \in \mathrm{Top}K(\bx) \,\right],
\end{gathered}
\end{equation}
where $\mathbf{1}[\cdot]$ is the indicator function. The weight $w_e \in [0,1]$ is the fraction of all routing decisions that select expert $e$ on the calibration dataset, so hot experts receive large weights and cold experts small ones, and $\sum_{e \in \mathcal{E}} w_e = 1$. This weighting ensures that the total MSE reflects the inference-time impact of each group: groups containing frequently activated experts contribute more to the overall approximation error.

This problem presents several challenges: $g$ is discrete and non-differentiable, $\mathrm{MSE}_m(g)$ can only be evaluated by training the united expert, and the search space is combinatorially large. Moreover, expert specialization differs across layers, necessitating per-layer optimization. These properties motivate a reinforcement learning approach. We apply policy optimization to expert grouping and solve it with GRPO, using the negative total MSE as the reward signal: $R(g) = -\mathrm{MSE}_{\text{total}}(g)$.

\subsection{GRPO-Based Expert Grouping Search}\label{sec:grpo}

\begin{figure*}[t]
\centering
\includegraphics[width=0.8\textwidth]{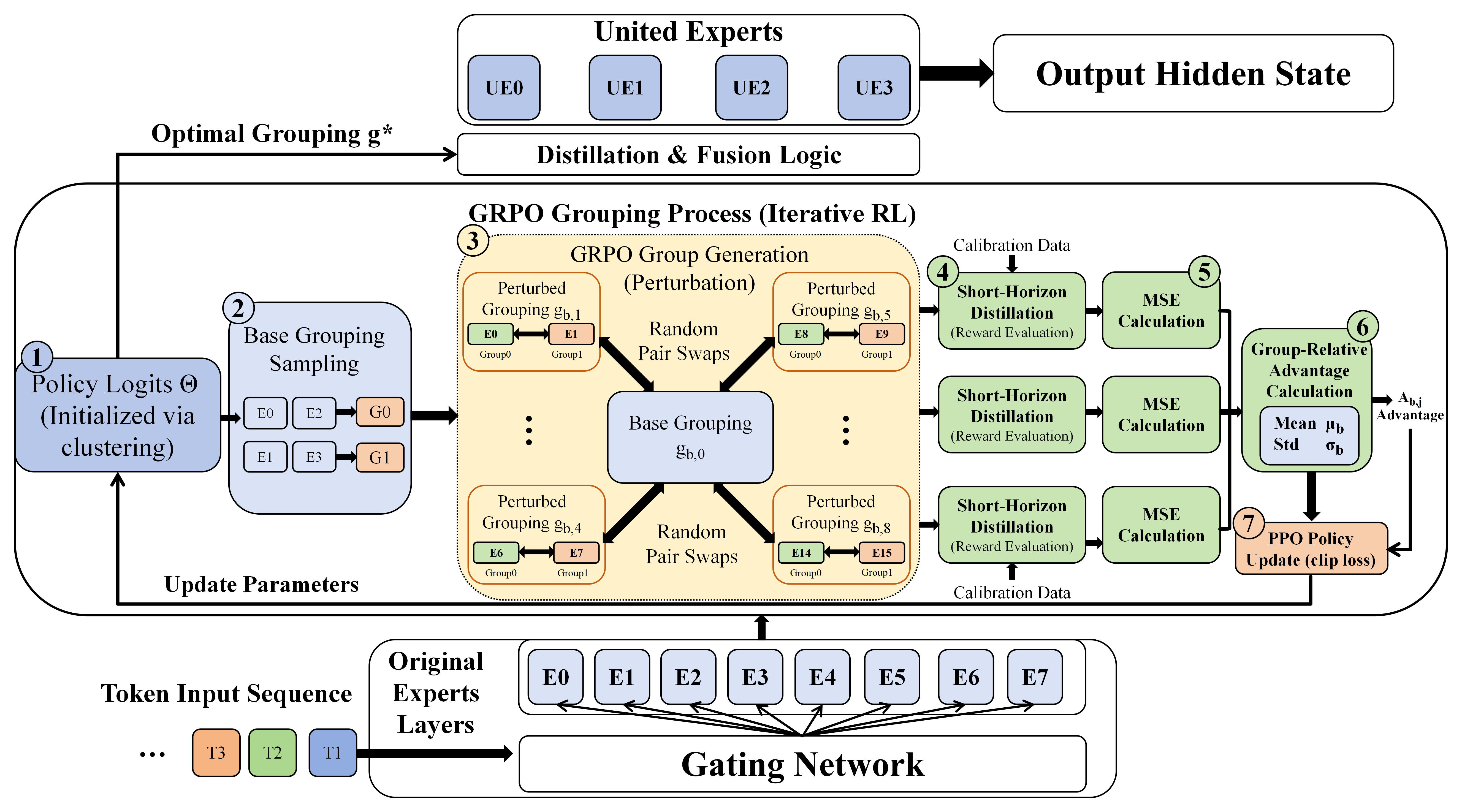}
\caption{GRPO-based expert grouping search process}
\label{fig:grpo_grouping}
\end{figure*}

\begin{algorithm*}[t]
\caption{GRPO-based expert grouping search}
\label{alg:grpo_search}
\begin{algorithmic}[1]
\State \textbf{Input:} Policy logits $\bTheta \in \mathbb{R}^{E \times G}$; calibration data $\mathcal{D}$; routing weights $w$; expert weights $\{\bW_e\}$; GRPO steps $N$; batch size $B$; perturbations per base $P$; swap count $n_{\text{swap}}$; distill steps $T_d$; PPO epochs $K_{\mathrm{PPO}}$; stabilizer $\epsilon_A$; clip $\epsilon_{\mathrm{clip}}$
\State \textbf{Output:} Optimal expert grouping $g^*$
\State $\bTheta_{\text{ref}} \leftarrow \bTheta;\quad R_{\text{best}} \leftarrow -\infty$
\For{$\text{step} = 1$ to $N$}
    \State $\bTheta_{\text{old}} \leftarrow \bTheta$
    \State \textbf{// Phase 1: Sample \& Perturb}
    \For{$b = 1$ to $B$}
        \State Sample base grouping $g_{b,0} \sim \pi_{\bTheta_{\text{old}}}$ with capacity constraints in Equation~(\ref{eq:policy})
        \For{$j = 1$ to $P$}
            \State $g_{b,j} \leftarrow \mathrm{SwapRandomPairs}(g_{b,0}, n_{\text{swap}})$
        \EndFor
    \EndFor
    \State \textbf{// Phase 2: Reward Evaluation}
    \For{each $g_{b,j}$, $b \in [1,B]$, $j \in [0,P]$}
        \State Run $T_d$-step distillation $\rightarrow \mathrm{MSE}_{\text{total}}(g_{b,j})$ via Equation~(\ref{eq:total_mse})
        \State $R(g_{b,j}) \leftarrow -\mathrm{MSE}_{\text{total}}(g_{b,j})$; store $\log \pi_{\bTheta_{\text{old}}}(g_{b,j})$
        \If{$R(g_{b,j}) > R_{\mathrm{best}}$}
            \State $R_{\mathrm{best}} \leftarrow R(g_{b,j});\quad g^* \leftarrow g_{b,j}$
        \EndIf
    \EndFor
    \State \textbf{// Phase 3: Group-Relative Advantage}
    \For{$b = 1$ to $B$}
        \State $\mu_b \leftarrow \frac{1}{P+1}\sum_{j=0}^{P} R(g_{b,j});\quad \sigma_b \leftarrow \mathrm{std}(\{R(g_{b,j})\}_{j=0}^{P})$
        \State $\hat{A}_{b,j} \leftarrow (R(g_{b,j}) - \mu_b) / (\sigma_b + \epsilon_A)$ for $j = 0, \ldots, P$
    \EndFor
    \State \textbf{// Phase 4: PPO Clipped Surrogate Update}
    \For{$\text{epoch} = 1$ to $K_{\mathrm{PPO}}$}
        \State $r_{b,j} \leftarrow \pi_{\bTheta}(g_{b,j}) / \pi_{\bTheta_{\text{old}}}(g_{b,j})$; $\mathcal{L}_{\text{GRPO}}$ via Equation~(\ref{eq:grpo_loss})
        \State Update $\bTheta$ via Adam; clip $\|\nabla \bTheta\| \leq 1.0$
    \EndFor
\EndFor
\State $g_{\text{greedy}} \leftarrow \arg\max_g \pi_{\bTheta}(g)$; evaluate $R(g_{\text{greedy}})$
\State \textbf{if} $R(g_{\text{greedy}}) > R_{\text{best}}$ \textbf{then} $g^* \leftarrow g_{\text{greedy}}$
\State \Return $g^*$
\end{algorithmic}
\end{algorithm*}

As illustrated in Figure~\ref{fig:grpo_grouping}, the grouping search follows an iterative loop. In each iteration, the current policy samples base groupings, which are perturbed by swapping expert pairs across groups to generate candidate variants. All candidates are evaluated through short-horizon distillation, where a lightweight student approximates the routing-weighted output of member experts and the resulting MSE serves as the quality metric. The group-relative advantage compares each candidate against its perturbation group, providing a variance-reduced gradient signal. The policy is then updated via clipped surrogate optimization, steering toward lower distillation error. This repeats until convergence or early stopping.

In this work, we employ GRPO to solve the expert grouping problem. The grouping policy is parameterized by a learnable logit matrix $\bTheta \in \mathbb{R}^{E \times G}$, where $\Theta_{e,m}$ encodes the affinity of expert $e$ for group $m$. Assignments are drawn sequentially from expert $0$ to $E{-}1$. At each step, the policy outputs a categorical distribution over available groups with capacity constraints:
\begin{equation}
\label{eq:policy}
\pi_{\bTheta}(g_e = m \mid e) = \frac{\exp(\Theta_{e,m})}{\sum_{m' \in \mathcal{A}_e} \exp(\Theta_{e,m'})},
\end{equation}
where $\mathcal{A}_e$ is the set of groups whose current member count has not yet reached the capacity $\mathrm{cap} = \lceil E/G \rceil$. The probability of a complete grouping is the product of per-expert assignment probabilities: $\pi_{\bTheta}(g) = \prod_{e} \pi_{\bTheta}(g_e \mid e)$.

The reward is defined as the negative total post-distillation MSE after short-horizon distillation. For each group $m$ in a sampled grouping $g$, the distillation target is the routing-weighted average of member expert outputs:
\begin{equation}
\label{eq:target}
\alpha_e = \frac{w_e}{\sum_{e' \in G_m} w_{e'}}, \qquad t_m(\bx) = \sum_{e \in G_m} \alpha_e\, f_e(\bx),
\end{equation}
where $w_e$ is the routing frequency of expert $e$ defined in Equation~(\ref{eq:we}). The united expert is initialized as the parameter average of member experts and optimized for $T_d$ steps with AdamW. The layer-level total MSE is:
\begin{equation}
\label{eq:total_mse}
\begin{gathered}
\beta_m = \sum_{e \in G_m} w_e,\\
\mathrm{MSE}_{\text{total}}(g) = \sum_{m=1}^{G} \beta_m \,\mathrm{MSE}_m,
\end{gathered}
\end{equation}
and the reward is $R(g) = -\mathrm{MSE}_{\text{total}}(g)$.

The key idea of GRPO is to compute advantages \emph{within groups} of related samples rather than across the entire batch. At each training step, we sample $B$ base groupings from the current policy, and for each base generate $P$ perturbations by swapping expert pairs across groups. Each base with its perturbations forms a GRPO group of size $P{+}1$. Since perturbed groupings share most expert assignments and differ only in swapped pairs, computing advantages within such correlated groups reduces gradient variance compared to batch-level normalization. The group-relative advantage for the $j$-th member of the $b$-th group is:
\begin{equation}
\label{eq:advantage}
\begin{gathered}
\hat{A}_{b,j} = \frac{R(g_{b,j}) - \mu_b}{\sigma_b + \epsilon_A}, \quad \mu_b = \frac{1}{P{+}1}\sum_{j=0}^{P} R(g_{b,j}),\\
\sigma_b = \mathrm{std}\big(\{R(g_{b,j})\}_{j=0}^{P}\big),
\end{gathered}
\end{equation}
where $\mu_b$ and $\sigma_b$ are computed \emph{within} the $b$-th group only, and $\epsilon_A>0$ prevents division by zero. Thus, each base is compared against its own perturbations.

Following GRPO, we use a Proximal Policy Optimization (PPO)-style clipped surrogate objective for stable updates over discrete grouping actions. The loss combines PPO clip-style policy updates with an entropy bonus and a Kullback--Leibler (KL) divergence penalty against the reference policy $\pi_{\text{ref}}$:
\begin{equation}
\label{eq:grpo_loss}
\begin{aligned}
\mathcal{L}_{\text{GRPO}} = {}&-\frac{1}{B(P{+}1)}\sum_{b=1}^{B}\sum_{j=0}^{P} L^{\text{clip}}_{b,j}\\
&- c_{\text{ent}}\,\bar{H} + c_{\text{kl}}\,\overline{D_{\text{KL}}},
\end{aligned}
\end{equation}
where
\begin{equation*}
\begin{aligned}
L^{\text{clip}}_{b,j} = \min\!\big(&r_{b,j} \hat{A}_{b,j},\\
&\mathrm{clip}(r_{b,j}, 1{-}\epsilon_{\mathrm{clip}}, 1{+}\epsilon_{\mathrm{clip}})\hat{A}_{b,j}\big),
\end{aligned}
\end{equation*}
and $r_{b,j} = \pi_{\bTheta}(g_{b,j}) / \pi_{\bTheta_{\text{old}}}(g_{b,j})$ is the importance ratio, $\bar{H}$ is the mean entropy, and $\overline{D_{\text{KL}}}$ is the mean KL divergence to $\pi_{\text{ref}}$. The policy is updated for $K_{\mathrm{PPO}}$ epochs per step with gradient clipping. The subscripts distinguish the advantage stabilizer $\epsilon_A$ from the PPO clipping range $\epsilon_{\mathrm{clip}}$.

In practice, we initialize $\bTheta$ using hierarchical clustering on the pairwise cosine similarity matrix of expert outputs from calibration data. This warm-start captures expert behavioral similarity, reducing the number of GRPO steps needed to converge. The clustering-derived assignment sets higher logit values for expert-group pairs placed together by clustering, with remaining logits initialized to zero. We further employ early stopping: if $R_{\text{best}}$ does not improve for a configurable number of consecutive steps, the search terminates early.

The complete search procedure is described in Algorithm~\ref{alg:grpo_search}, which consists of four phases: sampling and perturbation (lines~6--12), reward evaluation through short-horizon distillation (lines~13--20), group-relative advantage computation (lines~21--25), and PPO clipped surrogate update (lines~26--31).
Several implementation choices are made to connect the search objective with practical serving. Capacity-constrained sampling guarantees that every candidate grouping can be deployed without changing the expert-parallel layout. Swap-based perturbation creates local alternatives around the same base grouping, so the relative advantage compares candidates under similar structural contexts rather than unrelated partitions. The routing-weighted reward further prevents rarely activated experts from dominating the search objective. After the offline search finishes, only the selected grouping map and the distilled united expert weights are loaded by the serving engine, so GRPO introduces no additional request-time computation.
BrownoutMoE therefore treats GRPO as an offline calibration component rather than a request-path optimizer. The calibration run records both routing frequencies and expert outputs for each layer, and the final grouping is stored as a model-version artifact together with the distilled united-expert weights. When the workload changes, this artifact can be regenerated without modifying the online scheduler. During serving, the original gate still determines the active experts for each token; BrownoutMoE only replaces eligible cold-expert calls with their corresponding united expert according to the saved map. This separation keeps the online SLO controller simple: it adjusts the brownout threshold, while the structure-aware map determines the lower-cost substitute selected at that threshold.

The search objective is deliberately evaluated after a student has been fitted, rather than from expert-parameter distance alone. A candidate receives a high reward only when its members produce a target that one united expert can reproduce under the calibration distribution. Routing-frequency weights then give heavily used groups a proportionate influence on the decision. Capacity-constrained sampling keeps the number of members per group comparable and preserves the original expert-parallel layout. Together, these choices align the discrete policy update with the artifact that will actually be loaded by the serving engine.

The search is performed layer by layer, factorizing the joint assignment into smaller constrained decisions. This makes policy search tractable and preserves the parallel layout, but excludes cross-layer compensation effects from the short-horizon reward. The reward therefore ranks candidate maps rather than standing in for downstream benchmarks. The selected map is subjected to full united-expert distillation and end-to-end evaluation, with calibration traces kept separate from the reported task and serving workloads. Substantial hidden-state or routing shift should trigger regeneration of the model-version artifact because the online controller cannot correct a structural mismatch. These safeguards keep the offline search interpretable: it proposes a deployable grouping, and the runtime controller changes only how often that grouping is used.

\textbf{Complexity analysis.} The overall time complexity is dominated by Phase 2 (reward evaluation), requiring $\mathcal{O}(N \cdot B(P{+}1) \cdot T_d)$ distillation steps over $N$ GRPO iterations. Candidate storage and group-relative statistics require $\mathcal{O}(B(P{+}1))$ auxiliary memory in addition to model and optimizer states. In our implementation, reward evaluation accounts for over 95\% of wall-clock time, making the distillation budget $T_d$ the primary efficiency knob.

\subsection{Grouping-Consistent United Expert Distillation}\label{sec:distill}

After GRPO search finds the optimal grouping $g^*$ for each layer, we perform full distillation to train the united expert parameters:
\begin{equation}
\label{eq:full_distill}
\min_{\bW^{(m)}} \mathbb{E}_{\bx \sim \mathcal{D}} \left[ \| f_{u_m}(\bx; \bW^{(m)}) - t_m(\bx) \|_2^2 \right],
\end{equation}
where $\mathcal{D}$ is the calibration dataset and $t_m(\bx)$ is defined in Equation~(\ref{eq:target}). The united expert is initialized as the parameter average of member experts and trained with AdamW (learning rate $= 10^{-4}$, max steps $= 2000$) with early stopping (patience $= 300$, min-delta $= 10^{-7}$).

The united expert weights and the grouping map are saved and loaded by the inference engine at runtime. The grouping map $\{g^*(e) : e \in \mathcal{E}\}$ determines which united expert handles tokens originally routed to each cold expert, replacing the fixed sequential grouping.

\subsection{SLO-Aware Latency Control}\label{sec:slo}

BrownoutMoE uses an SLO-aware latency controller to decide how aggressively the runtime applies united experts. The grouping map determines which united expert serves each cold expert, while threshold adjustment determines when to use that path. The controller maximizes accuracy subject to a per-token latency objective:
\begin{equation}
\label{eq:slo}
\text{maximize}_{\tau \in [0,1]} \quad Q(\tau,k)
\end{equation}
\begin{equation}
\label{eq:slo_constraint}
\text{subject to} \quad \ell_{90}(\tau,k) \leq s,
\end{equation}
where $Q(\tau,k)$ is the expected answer-quality score under threshold $\tau$ and grouping factor $k$, and $\ell_{90}(\tau,k)$ is the resulting P90 latency. The controller therefore changes only $\tau$ at runtime while the grouping factor and united-expert weights remain fixed for a model version.

\begin{algorithm*}[t]
\caption{SLO-aware latency control}
\label{alg:slo_control}
\begin{algorithmic}[1]
\State \textbf{Input:} Initial threshold $\tau_0$; target SLO $s$; warning factor $\rho$; control horizon $H$; window stream $\{W_t\}$; load thresholds $q_{\mathrm{low}},q_{\mathrm{high}}$; increment $\Delta$; shrink ratio $\gamma$
\State \textbf{Output:} Threshold sequence $\{\tau_t\}_{t=1}^{H}$
\State $\ell_{\mathrm{warn}} \leftarrow \rho s$
\For{$t = 0$ to $H-1$}
    \State $\ell_{90,t} \leftarrow \operatorname{P90Latency}(W_t)$; $q_t \leftarrow \operatorname{Load}(W_t)$
    \If{$\ell_{90,t} > s$ \textbf{or} $q_t \geq q_{\mathrm{high}}$} \Comment{high-load protection}
        \State $\tau_{t+1} \leftarrow \max(0,\gamma\tau_t)$
    \ElsIf{$\ell_{90,t} < \ell_{\mathrm{warn}}$ \textbf{and} $q_t \leq q_{\mathrm{low}}$} \Comment{low-load recovery}
        \State $\tau_{t+1} \leftarrow \min(1,\tau_t + \Delta)$
    \Else
        \State $\tau_{t+1} \leftarrow \tau_t$
    \EndIf
    \State Apply $\tau_{t+1}$ to requests arriving in interval $t+1$
\EndFor
\State \Return $\{\tau_t\}_{t=1}^{H}$
\end{algorithmic}
\end{algorithm*}

Let $\tau_t \in [0,1]$ denote the threshold at control interval $t$, let $s$ denote the target SLO, and let $W_t$ contain the latency observations collected during the recent window. We define a warning line below the SLO and use the recent P90 latency as feedback:
\begin{equation}
\label{eq:slo_warning}
\begin{gathered}
\ell_{\mathrm{warn}} = \rho s, \qquad 0 < \rho < 1,\\
\ell_{90,t} = \operatorname{P90Latency}(W_t).
\end{gathered}
\end{equation}
Let $q_t \in [0,1]$ be a normalized load signal obtained from the request arrival rate and queue occupancy, and choose two load thresholds $q_{\mathrm{low}} < q_{\mathrm{high}}$. The controller treats an interval as overloaded when either the P90 latency exceeds the SLO or $q_t$ exceeds $q_{\mathrm{high}}$. It treats an interval as lightly loaded only when both the P90 latency is below the warning line and $q_t$ is below $q_{\mathrm{low}}$. This two-signal rule prevents a short latency dip from immediately restoring optional computation while the queue is still growing. The update has a dead band between the warning line and the SLO; within that band the current mode is held. The bounds in Equation~(\ref{eq:slo_update}) keep the threshold valid after repeated updates:
\begin{equation}
\label{eq:slo_update}
\tau_{t+1} =
\begin{cases}
\max(0,\gamma\tau_t), & \ell_{90,t} > s \text{ or } q_t \geq q_{\mathrm{high}},\\
\min(1,\tau_t + \Delta), & \ell_{90,t} < \ell_{\mathrm{warn}} \text{ and}\\
 & \quad q_t \leq q_{\mathrm{low}},\\
\tau_t, & \text{otherwise},
\end{cases}
\end{equation}
where $\Delta$ is the recovery increment and $\gamma \in (0,1)$ is the overload shrinkage ratio. A decreasing threshold makes more cold-expert calls eligible for united experts, reducing optional computation and protecting the tail. An increasing threshold restores original-expert computation and therefore recovers answer quality when load is low. This feedback design follows the principle that tail latency, rather than mean latency, controls interactive-service quality~\cite{dean2013tail}. It is also consistent with SLO-targeted resource controllers that adjust optional work as demand changes~\cite{wang2024autothrottle}.

\textbf{Algorithm design.} Algorithm~\ref{alg:slo_control} summarizes the repeated control loop. Lines~3--5 initialize the warning line and refresh the latency and load signals; lines~6--11 apply the high-load, low-load, or dead-band update. Line~13 applies the new threshold to subsequent requests, and line~14 advances to the next interval. Because the update is outside the model forward pass, it adds no per-token overhead and applies unchanged to zero-, partial-, and full-brownout configurations.

\textbf{Complexity analysis.} Let $n$ be the number of observations in each monitoring window and $H$ the number of control intervals. Exact P90 computation costs $\mathcal{O}(n\log n)$ per interval, while load aggregation and threshold update cost $\mathcal{O}(n)$ and $\mathcal{O}(1)$. Thus, Algorithm~\ref{alg:slo_control} takes $\mathcal{O}(Hn\log n)$ time and $\mathcal{O}(n+H)$ space when the threshold sequence is stored, or $\mathcal{O}(n)$ streaming space. The controller is independent of model size and adds no per-token inference cost.

\section{Performance Evaluation}\label{sec:eval}

This section evaluates BrownoutMoE from three complementary perspectives: downstream quality under increasingly aggressive grouping, serving throughput under fused and unfused MoE execution, and scalability and latency under different GPU counts and request rates. In all experiments, the checkpoint, router, calibration traces, batching policy, and SLO controller remain fixed for all grouping methods; the controlled variable is the map from original experts to united experts, which is the only artifact that changes between BrownoutServe and BrownoutMoE. This scope separates offline grouping quality from online execution: BrownoutMoE learns a behavior-validated substitution map while retaining the original gate and batching policy. The map is tied to a checkpoint and calibration distribution, whereas online control selects only the substitution level.

\subsection{Experimental Setup}\label{sec:setup}

\textbf{Testbed and model.} We evaluate Qwen1.5-MoE-A2.7B-Chat, a 14.3B-parameter MoE model with 60 routed experts per layer and top-4 routing~\cite{bai2023qwen}. Experiments use eight NVIDIA A100-80GB GPUs, with 24 MoE layers executed in parallel.

\textbf{Datasets.} We use widely used datasets to evaluate the performance of our proposed approach:

\emph{PIQA.} The Physical Interaction: Question Answering benchmark tests commonsense knowledge about everyday physical situations through two-choice questions~\cite{bisk2020piqa}. It is useful for this study because small changes in the model's functional behavior can change the preferred physical explanation, making it sensitive to approximation error.

\emph{COPA.} The Choice of Plausible Alternatives benchmark evaluates causal reasoning: given a premise, the model chooses which of two alternatives is the more plausible cause or effect~\cite{gordon2012copa}. Its compact two-choice format makes accuracy changes easy to compare across grouping factors while still probing a distinct reasoning skill from PIQA.

\emph{C-Eval.} C-Eval is a Chinese multi-subject knowledge benchmark that covers humanities, science, social science, and professional subjects~\cite{huang2023ceval}. We use its validation split with few-shot prompts. The broad subject coverage tests whether a united expert preserves behavior across heterogeneous knowledge domains rather than overfitting to one task type.

\emph{OpenBookQA.} OpenBookQA (OBQA) combines a small set of elementary science facts with multi-step commonsense reasoning questions~\cite{mihaylov2018openbookqa}. We use five-shot prompting. Compared with PIQA and COPA, it places more weight on combining stored facts with an inference chain, providing a complementary quality test.

\emph{Serving traces.} ShareGPT and Alpaca are conversational instruction-following traces used for throughput and latency experiments. Their prompt and response lengths differ, so they expose the serving system to distinct batching and queueing patterns. We follow the request-generation and measurement conventions of the vLLM evaluation protocol~\cite{kwon2023efficient}.

\textbf{Baselines.} We compare our proposed approach with several baselines.

\emph{BrownoutServe (Sequential Grouping).} BrownoutServe~\cite{11357542} is the original system with sequential expert grouping: experts $0, \ldots, k{-}1$ form group 0, the next $k$ experts form group 1, and so on. United experts are trained by knowledge distillation under this fixed map, and partial brownout routes eligible cold-expert calls to the united path according to a threshold. This baseline isolates the benefit of replacing index-order grouping with behavior-aware grouping. It also represents a non-learning, greedy grouping strategy: the map is constructed directly from expert indices without any behavioral signal or search, so the comparison against it isolates the benefit of the GRPO-based search over a greedy grouping rule.

\emph{Zero Brownout.} Zero Brownout uses threshold $\tau=1.0$ and always keeps the original expert path. It provides the quality reference because no united-expert substitution is applied, while its throughput represents the cost of executing the complete original expert inventory.

\emph{Full Brownout.} Full Brownout uses threshold $\tau=0$ and sends all eligible cold-expert calls to united experts. It represents the most aggressive efficiency setting and serves as a lower quality reference for the accuracy plots.

Accuracy is reported as the percentage of correct answers. Serving performance is reported as generated tokens per second, throughput versus request rate, and latency under the same batching and GPU configuration.

\subsection{Accuracy Comparison}\label{sec:accuracy}

We separate the downstream accuracy results into three figures so that each grouping factor can be inspected independently. Every figure contains four horizontally aligned panels for PIQA, COPA, C-Eval, and OBQA, using the same baseline legend and brownout-threshold sweep.

\begin{figure*}[t]
\centering
\begin{tikzpicture}[x=0.48cm, y=0.48cm]
\definecolor{cseq}{HTML}{4472C4}
\definecolor{cgrpo}{HTML}{ED7D31}
\definecolor{cfb}{HTML}{FF0000}
\draw[color=cgrpo, line width=0.8pt] (0,0) -- (1.2,0);
\fill[cgrpo] (0.6,0) ++(-1.8pt,-1.8pt) rectangle ++(3.6pt,3.6pt);
\node[right, font=\scriptsize] at (1.4,0) {BrownoutMoE};
\draw[color=cseq, line width=0.8pt] (6.0,0) -- (7.2,0);
\fill[cseq] (6.6,0) circle (1.8pt);
\node[right, font=\scriptsize] at (7.4,0) {BrownoutServe};
\draw[color=cfb, line width=0.8pt] (12.6,0) -- (13.8,0);
\fill[cfb] (13.2,0) -- ++(2pt,2.8pt) -- ++(2pt,-2.8pt) -- ++(-4pt,0pt) -- cycle;
\node[right, font=\scriptsize] at (14.0,0) {Full Brownout};
\draw[black, dashed, line width=0.7pt] (19.2,0) -- (20.4,0);
\node[right, font=\scriptsize] at (20.6,0) {Zero Brownout};
\end{tikzpicture}

\vspace{2pt}
\includegraphics[width=0.247\textwidth]{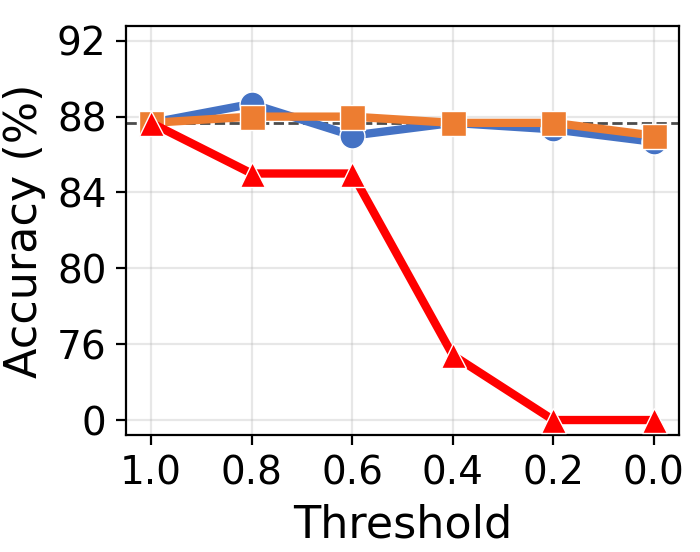}%
\hspace{0.002\textwidth}%
\includegraphics[width=0.247\textwidth]{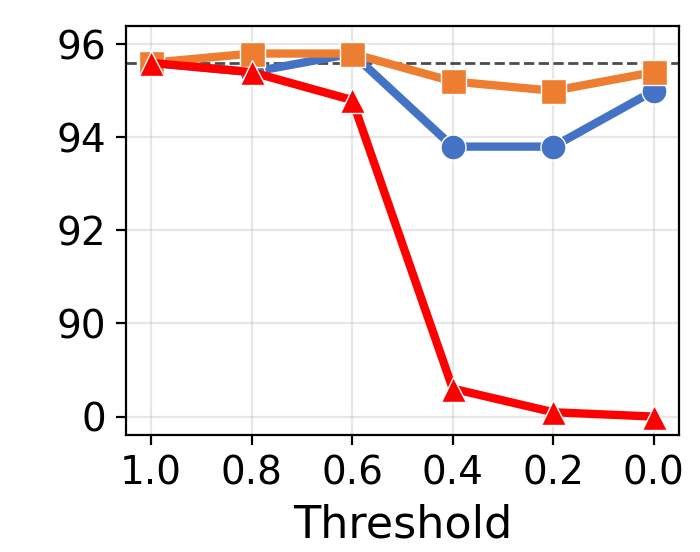}%
\hspace{0.002\textwidth}%
\includegraphics[width=0.247\textwidth]{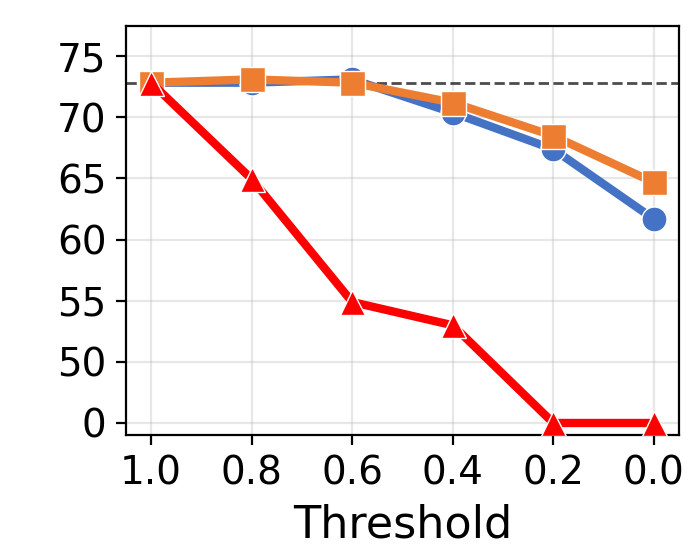}%
\hspace{0.002\textwidth}%
\includegraphics[width=0.247\textwidth]{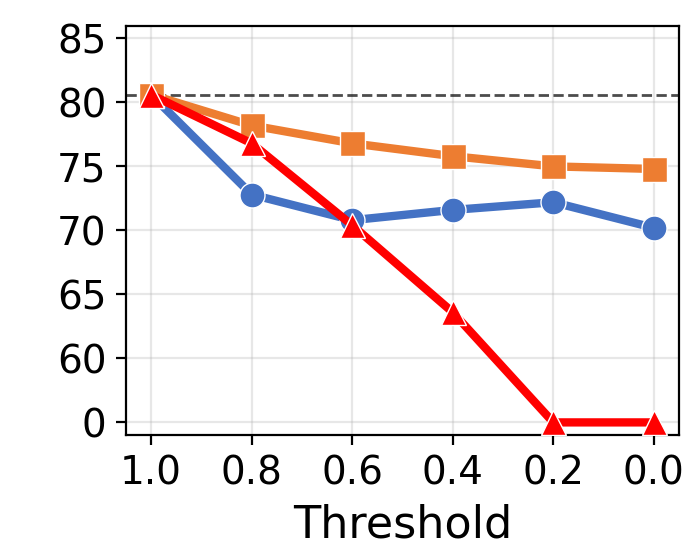}

\vspace{2pt}
\noindent\makebox[0.249\textwidth][c]{\small (a) PIQA}%
\makebox[0.249\textwidth][c]{\small (b) COPA}%
\makebox[0.249\textwidth][c]{\small (c) C-Eval}%
\makebox[0.249\textwidth][c]{\small (d) OBQA}
\caption{Accuracy versus brownout threshold under 2-way grouping on (a) PIQA, (b) COPA, (c) C-Eval, and (d) OBQA}
\label{fig:acc_2way}
\end{figure*}

\textbf{2-way grouping.} Figure~\ref{fig:acc_2way} reports the least aggressive setting, in which each united expert represents two original experts. Zero Brownout, shown by the dashed line, retains the original expert path and therefore provides fixed quality references of approximately 87.6\% on PIQA, 95.6\% on COPA, 72.8\% on C-Eval, and 80.5\% on OBQA. Under BrownoutServe's sequential grouping, PIQA remains close to the reference but decreases from about 87.6\% to 86.6\% as the threshold changes from 1.0 to 0.0. COPA is less monotonic: its accuracy falls to roughly 93.8\% at thresholds 0.4 and 0.2 before recovering to about 95.0\% at 0.0. The more structure-sensitive datasets show larger losses. C-Eval decreases steadily to approximately 61.5\%, while OBQA drops sharply at threshold 0.8 and finishes near 70.0\%. BrownoutMoE follows the same workload-dependent trend but remains more stable: its endpoint accuracies are about 87.0\%, 95.4\%, 64.5\%, and 74.8\% on the four datasets, respectively. The moderate operating point at threshold 0.4 shows the same pattern before the curves reach their endpoints: BrownoutMoE is approximately 0.3\%, 1.4\%, 1\%, and 4\% higher on PIQA, COPA, C-Eval, and OBQA, respectively. This comparison is important because it shows that the learned grouping improves the usable quality--efficiency region rather than helping only under full substitution. At threshold 0.0, the learned grouping still recovers roughly 0.4\% on PIQA, 0.4\% on COPA, 3\% on C-Eval, and nearly 5\% on OBQA compared with sequential grouping. Full Brownout is the most aggressive scheme and produces the steepest degradation after threshold 0.6, confirming that replacing all eligible expert calls without controlling substitution intensity is unsuitable for quality-sensitive requests. Overall, the comparatively small BrownoutMoE--BrownoutServe gap on PIQA and COPA shows that many two-expert groups are still easy to distill, whereas the clearer C-Eval and OBQA margins demonstrate that behavior-aware pairing is useful even at the smallest grouping factor.

Across the complete threshold sweep, BrownoutMoE narrows the observed accuracy range from roughly 11.3\% to 8.3\% on C-Eval and from about 10.5\% to 5.7\% on OBQA. Zero Brownout therefore supplies the fixed quality ceiling but no expert reduction, Full Brownout supplies the lower quality bound, and the two partial schemes expose how the grouping decision changes the usable operating region.

\begin{figure*}[t]
\centering
\begin{tikzpicture}[x=0.48cm, y=0.48cm]
\definecolor{cseq}{HTML}{4472C4}
\definecolor{cgrpo}{HTML}{ED7D31}
\definecolor{cfb}{HTML}{FF0000}
\draw[color=cgrpo, line width=0.8pt] (0,0) -- (1.2,0);
\fill[cgrpo] (0.6,0) ++(-1.8pt,-1.8pt) rectangle ++(3.6pt,3.6pt);
\node[right, font=\scriptsize] at (1.4,0) {BrownoutMoE};
\draw[color=cseq, line width=0.8pt] (6.0,0) -- (7.2,0);
\fill[cseq] (6.6,0) circle (1.8pt);
\node[right, font=\scriptsize] at (7.4,0) {BrownoutServe};
\draw[color=cfb, line width=0.8pt] (12.6,0) -- (13.8,0);
\fill[cfb] (13.2,0) -- ++(2pt,2.8pt) -- ++(2pt,-2.8pt) -- ++(-4pt,0pt) -- cycle;
\node[right, font=\scriptsize] at (14.0,0) {Full Brownout};
\draw[black, dashed, line width=0.7pt] (19.2,0) -- (20.4,0);
\node[right, font=\scriptsize] at (20.6,0) {Zero Brownout};
\end{tikzpicture}

\vspace{2pt}
\includegraphics[width=0.247\textwidth]{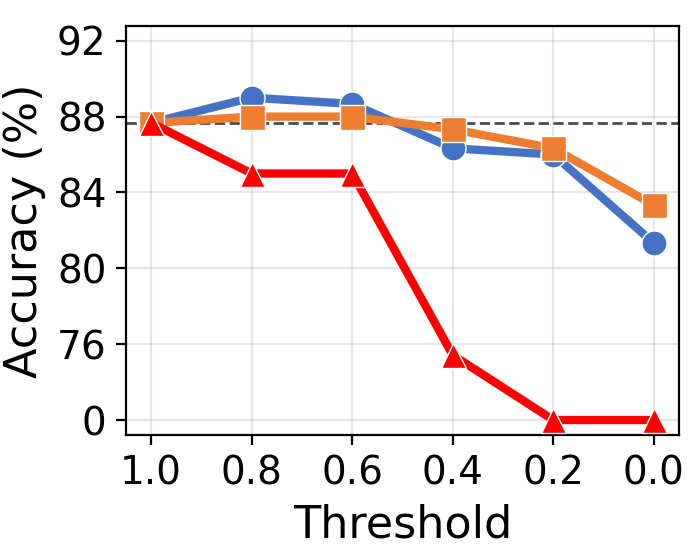}%
\hspace{0.002\textwidth}%
\includegraphics[width=0.247\textwidth]{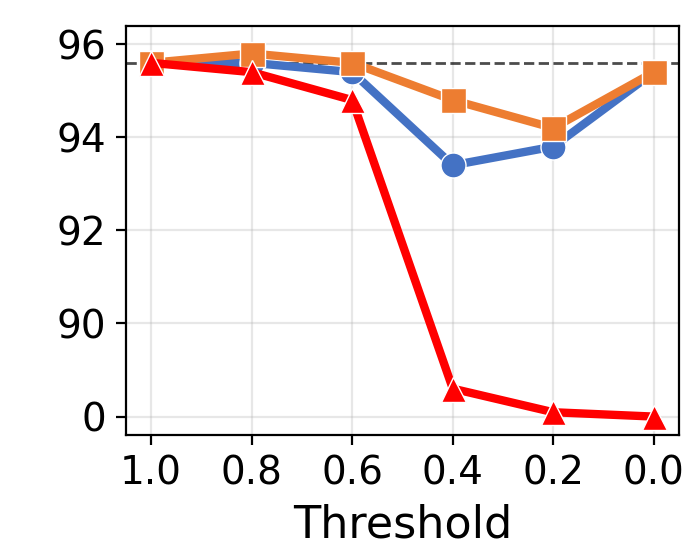}%
\hspace{0.002\textwidth}%
\includegraphics[width=0.247\textwidth]{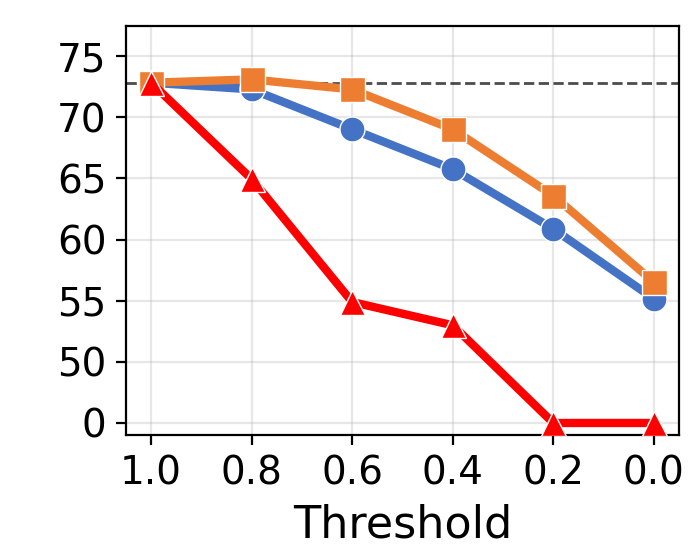}%
\hspace{0.002\textwidth}%
\includegraphics[width=0.247\textwidth]{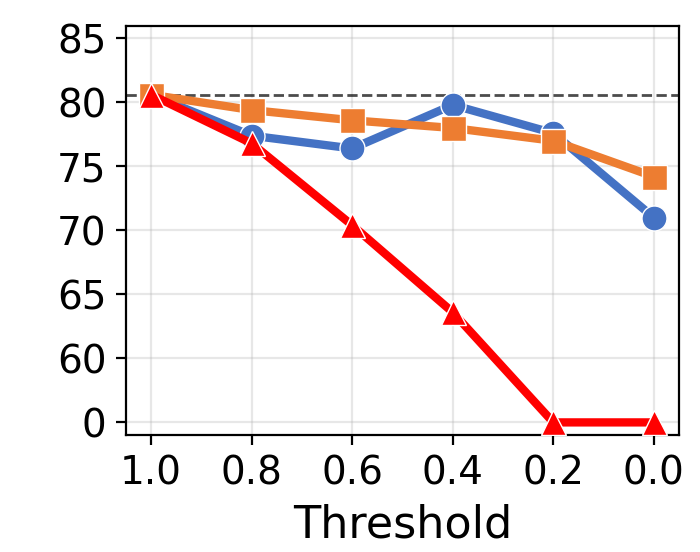}

\vspace{2pt}
\noindent\makebox[0.249\textwidth][c]{\small (a) PIQA}%
\makebox[0.249\textwidth][c]{\small (b) COPA}%
\makebox[0.249\textwidth][c]{\small (c) C-Eval}%
\makebox[0.249\textwidth][c]{\small (d) OBQA}
\caption{Accuracy versus brownout threshold under 4-way grouping on (a) PIQA, (b) COPA, (c) C-Eval, and (d) OBQA}
\label{fig:acc_4way}
\end{figure*}

\textbf{4-way grouping.} Figure~\ref{fig:acc_4way} increases the approximation burden because one united expert must now cover four original expert functions. The Zero Brownout references remain unchanged, which makes the additional loss attributable to the grouping and substitution decisions rather than to a different model or dataset. BrownoutServe preserves PIQA near 88\% at thresholds 1.0--0.6, but then declines to about 86.4\% at 0.4 and 81.2\% at 0.0. BrownoutMoE degrades more gradually and ends near 83.3\%, providing a gain of about 2.1\% at the most aggressive threshold. On COPA, both partial-brownout schemes remain above 93\%; however, BrownoutMoE stays around 94.8\% at threshold 0.4 compared with approximately 93.4\% for BrownoutServe, and both recover close to 95.4\% at threshold 0.0. C-Eval reveals the grouping effect most consistently: BrownoutServe falls from about 72.8\% to 65.8\%, 60.9\%, and 55.0\% at thresholds 0.4, 0.2, and 0.0, whereas BrownoutMoE retains roughly 69.0\%, 63.5\%, and 56.6\%. This corresponds to a reduction of up to 46.1\% in the degradation from the Zero Brownout reference under the reported setting. OBQA exhibits a non-monotonic BrownoutServe curve, briefly reaching about 79.8\% at threshold 0.4 before falling to 70.8\% at 0.0; BrownoutMoE is smoother and finishes near 74.2\%, about 3.4\% higher.

The Full Brownout curve provides a useful lower bound for interpreting these partial-brownout results. PIQA and COPA remain close to their original accuracies at thresholds 0.8 and 0.6, but both deteriorate sharply once the threshold reaches 0.4; the decline is even more pronounced on C-Eval and OBQA, whose Full Brownout curves fall to about 54.8\% and 70.0\%, respectively, by threshold 0.6 and continue downward thereafter. This behavior shows that the threshold controls more than a smooth performance knob: beyond a task-dependent operating point, indiscriminate united-expert use removes expert capacity required by the downstream task. Comparing the two partial schemes across the whole sweep, BrownoutServe occasionally matches or briefly exceeds BrownoutMoE at an isolated point, such as OBQA at threshold 0.4, but it does not maintain that advantage as substitution becomes more aggressive. BrownoutMoE instead provides the stronger overall envelope: it is never far from the better partial-brownout result on PIQA and COPA, while it establishes consistent margins on C-Eval and at the low-threshold end of OBQA.

Viewed scheme by scheme, Zero Brownout therefore defines the quality ceiling but offers no reduction in expert execution; Full Brownout maximizes substitution but crosses an unacceptable quality boundary on every dataset; BrownoutServe provides intermediate operating points but is sensitive to which experts happen to share an index-defined group; and BrownoutMoE produces a flatter, more predictable quality curve by grouping experts whose outputs can be distilled by the same student. The cross-dataset comparison also reveals different sensitivities: COPA tolerates 4-way substitution relatively well, PIQA begins to separate only at low thresholds, while C-Eval and OBQA expose grouping errors earlier. Around threshold 0.4, BrownoutMoE retains most of the Zero Brownout accuracy on PIQA and COPA and a measurable advantage on C-Eval, making this region a reasonable quality--efficiency compromise before online SLO feedback selects a more aggressive mode. These results position 4-way grouping as a practical middle ground: it removes substantially more expert work than 2-way grouping, while structure-aware grouping prevents the quality loss from growing at the same rate as sequential grouping. They also indicate that threshold selection should be dataset-aware, whereas the learned grouping supplies a more robust model artifact across those operating points.

At threshold 0.0, BrownoutServe retains approximately 92.7\%, 99.4\%, 75.5\%, and 88.0\% of the Zero Brownout accuracy on PIQA, COPA, C-Eval, and OBQA, respectively; BrownoutMoE raises the corresponding retention to about 95.1\%, 99.8\%, 77.7\%, and 92.2\%. At threshold 0.4, BrownoutMoE leads BrownoutServe by roughly 1\%, 1.4\%, and 3\% on PIQA, COPA, and C-Eval, respectively, although BrownoutServe briefly performs better on OBQA. This exception prevents overgeneralizing a single operating point and reinforces the need for the online controller to select thresholds from measured task behavior.

\begin{figure*}[t]
\centering
\begin{tikzpicture}[x=0.48cm, y=0.48cm]
\definecolor{cseq}{HTML}{4472C4}
\definecolor{cgrpo}{HTML}{ED7D31}
\definecolor{cfb}{HTML}{FF0000}
\draw[color=cgrpo, line width=0.8pt] (0,0) -- (1.2,0);
\fill[cgrpo] (0.6,0) ++(-1.8pt,-1.8pt) rectangle ++(3.6pt,3.6pt);
\node[right, font=\scriptsize] at (1.4,0) {BrownoutMoE};
\draw[color=cseq, line width=0.8pt] (6.0,0) -- (7.2,0);
\fill[cseq] (6.6,0) circle (1.8pt);
\node[right, font=\scriptsize] at (7.4,0) {BrownoutServe};
\draw[color=cfb, line width=0.8pt] (12.6,0) -- (13.8,0);
\fill[cfb] (13.2,0) -- ++(2pt,2.8pt) -- ++(2pt,-2.8pt) -- ++(-4pt,0pt) -- cycle;
\node[right, font=\scriptsize] at (14.0,0) {Full Brownout};
\draw[black, dashed, line width=0.7pt] (19.2,0) -- (20.4,0);
\node[right, font=\scriptsize] at (20.6,0) {Zero Brownout};
\end{tikzpicture}

\vspace{2pt}
\includegraphics[width=0.247\textwidth]{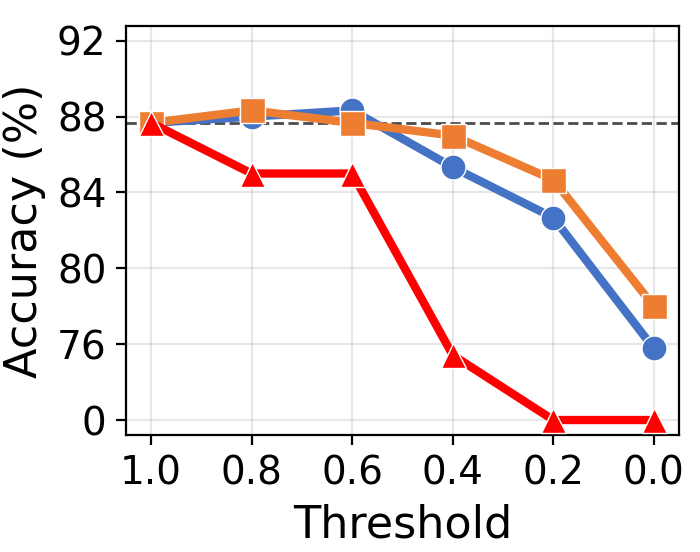}%
\hspace{0.002\textwidth}%
\includegraphics[width=0.247\textwidth]{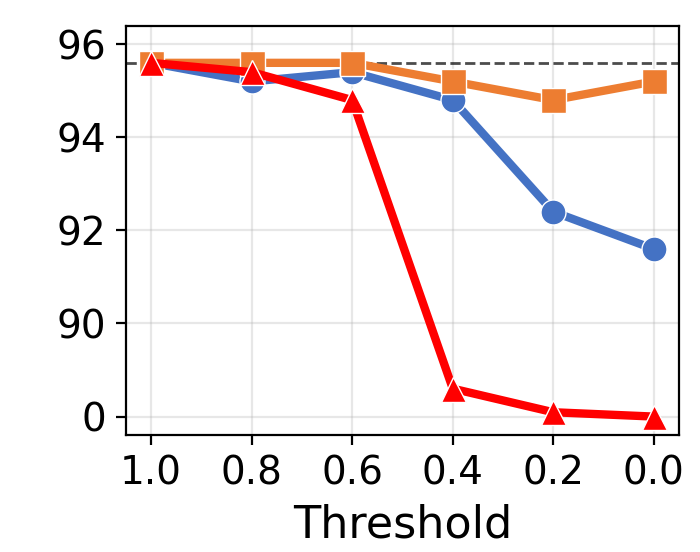}%
\hspace{0.002\textwidth}%
\includegraphics[width=0.247\textwidth]{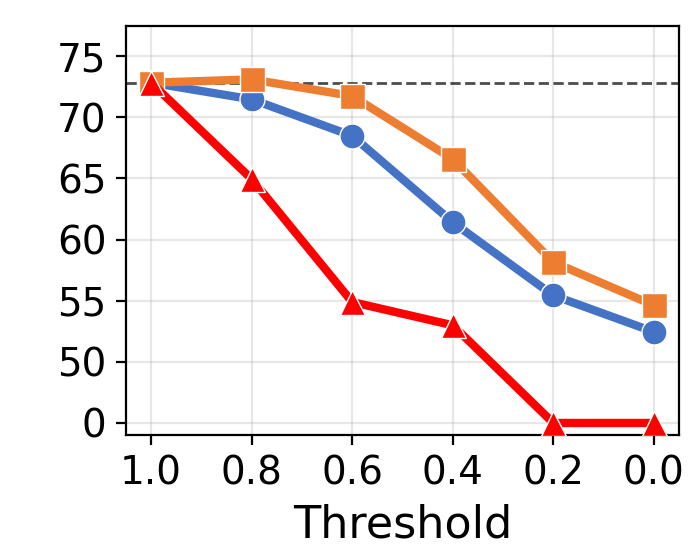}%
\hspace{0.002\textwidth}%
\includegraphics[width=0.247\textwidth]{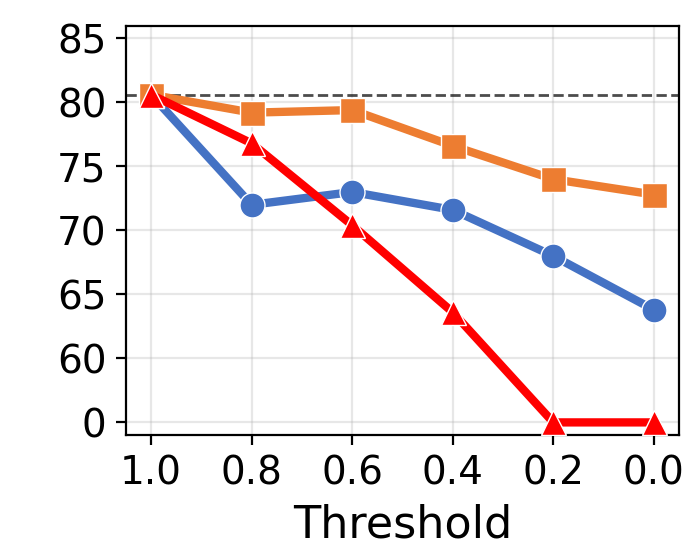}

\vspace{2pt}
\noindent\makebox[0.249\textwidth][c]{\small (a) PIQA}%
\makebox[0.249\textwidth][c]{\small (b) COPA}%
\makebox[0.249\textwidth][c]{\small (c) C-Eval}%
\makebox[0.249\textwidth][c]{\small (d) OBQA}
\caption{Accuracy versus brownout threshold under 8-way grouping on (a) PIQA, (b) COPA, (c) C-Eval, and (d) OBQA}
\label{fig:acc_8way}
\end{figure*}

\textbf{8-way grouping.} Figure~\ref{fig:acc_8way} presents the most demanding configuration, with eight original experts compressed into each united expert, and consequently produces the clearest separation between the two grouping strategies. Zero Brownout again fixes the original-model references at approximately 87.6\%, 95.6\%, 72.8\%, and 80.5\%. For PIQA, BrownoutServe declines to about 85.3\% at threshold 0.4, 82.6\% at 0.2, and 75.7\% at 0.0; BrownoutMoE preserves approximately 87.0\%, 84.5\%, and 78.0\% at the same points. On COPA, the BrownoutMoE curve remains close to 95\% throughout the sweep, while BrownoutServe drops more sharply after threshold 0.4 and reaches only about 91.6\% at 0.0. C-Eval shows a sustained advantage across the aggressive region: BrownoutMoE retains about 66.5\% at threshold 0.4 compared with 61.5\% for BrownoutServe, and remains about 2--3\% higher at thresholds 0.2 and 0.0. OBQA displays the largest absolute separation. BrownoutServe falls immediately to about 72\% at threshold 0.8, reaches about 71.5\% at 0.4, and ends near 64\%, whereas BrownoutMoE remains around 79\% at 0.8, 76.7\% at 0.4, and 72.8\% at 0.0. Quantitatively, at threshold 0.4 BrownoutMoE reduces accuracy degradation by 71.4\% on PIQA, 50.0\% on COPA, 45.3\% on C-Eval, and about 58\% on OBQA, where the reduction is computed as $(\Delta_{\text{seq}} - \Delta_{\text{ours}})/\Delta_{\text{seq}}$ with $\Delta_{\text{seq}}$ and $\Delta_{\text{ours}}$ denoting the accuracy drops of sequential grouping and BrownoutMoE from the Zero Brownout reference at the same threshold; OBQA also shows the largest absolute improvement over sequential grouping, about 7.2\% at threshold 0.8 and about 8.8\% at threshold 0.0. Full Brownout remains substantially below all partial-brownout alternatives, illustrating the quality risk of unconditional substitution. The widening advantage from 2-way to 8-way grouping verifies the central hypothesis: as one student covers more activation regions, index proximity becomes a poor proxy for functional compatibility, while GRPO-based grouping preserves complementary expert behaviors. Because this search and distillation are performed offline, the accuracy gain does not add computation to the request path.

\subsection{Throughput Comparison}\label{sec:throughput}

\begin{figure*}[t]
\centering
\begin{tikzpicture}[x=0.48cm, y=0.48cm]
\definecolor{cbase}{HTML}{4472C4}
\definecolor{c2w}{HTML}{ED7D31}
\definecolor{c4w}{HTML}{70AD47}
\definecolor{c8w}{HTML}{FF0000}
\draw[color=cbase, line width=0.8pt] (0,0) -- (1.2,0);
\fill[cbase] (0.6,0) circle (1.8pt);
\node[right, font=\scriptsize] at (1.4,0) {vLLM};
\draw[color=c2w, line width=0.8pt] (5.5,0) -- (6.7,0);
\fill[c2w] (6.1,0) ++(0pt,2.2pt) -- ++(2.2pt,-2.2pt) -- ++(-2.2pt,-2.2pt) -- ++(-2.2pt,2.2pt) -- cycle;
\node[right, font=\scriptsize] at (6.9,0) {2-way};
\draw[color=c4w, line width=0.8pt] (11.0,0) -- (12.2,0);
\fill[c4w] (11.6,0) -- ++(2pt,2.8pt) -- ++(2pt,-2.8pt) -- ++(-4pt,0pt) -- cycle;
\node[right, font=\scriptsize] at (12.4,0) {4-way};
\draw[color=c8w, line width=0.8pt] (16.5,0) -- (17.7,0);
\fill[c8w] (17.1,0) ++(-1.8pt,-1.8pt) rectangle ++(3.6pt,3.6pt);
\node[right, font=\scriptsize] at (17.9,0) {8-way};
\end{tikzpicture}

\vspace{2pt}
\includegraphics[width=0.247\textwidth]{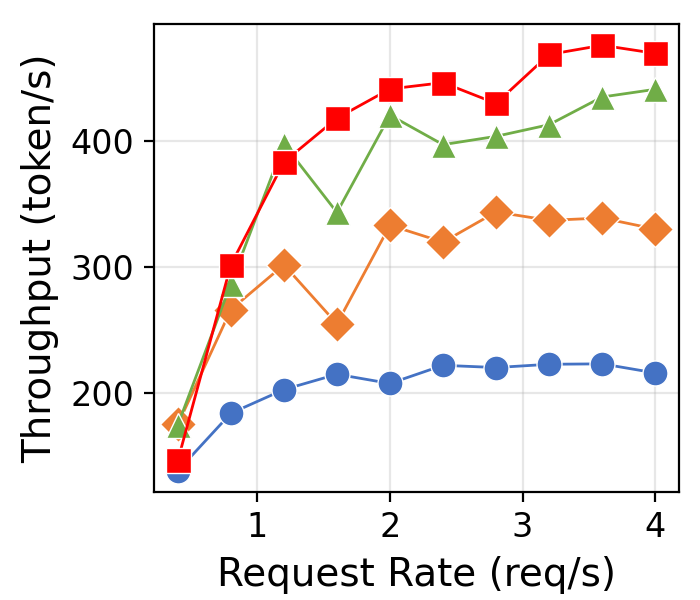}%
\hspace{0.002\textwidth}%
\includegraphics[width=0.247\textwidth]{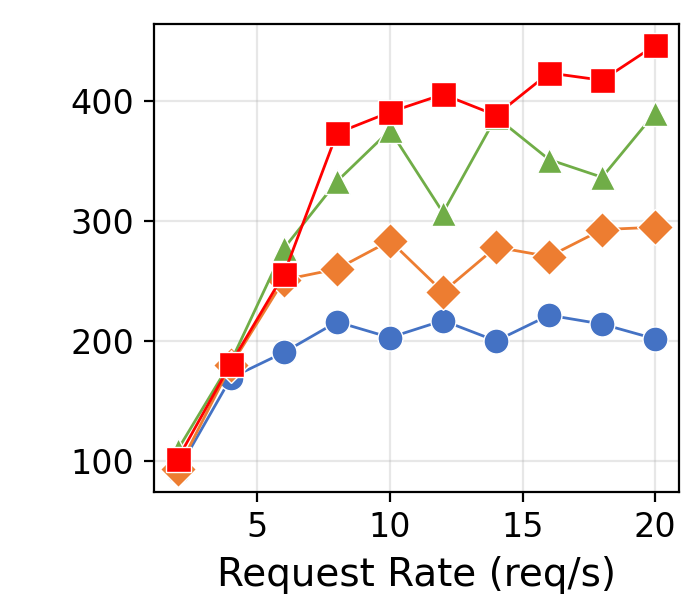}%
\hspace{0.002\textwidth}%
\includegraphics[width=0.247\textwidth]{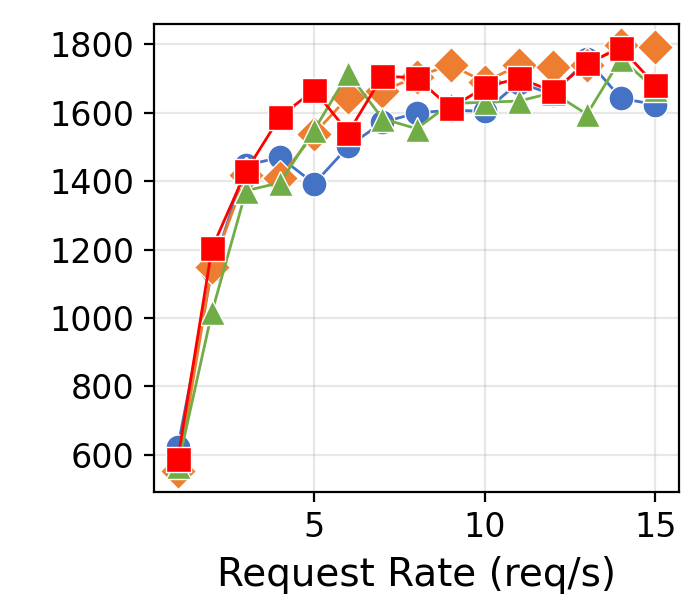}%
\hspace{0.002\textwidth}%
\includegraphics[width=0.247\textwidth]{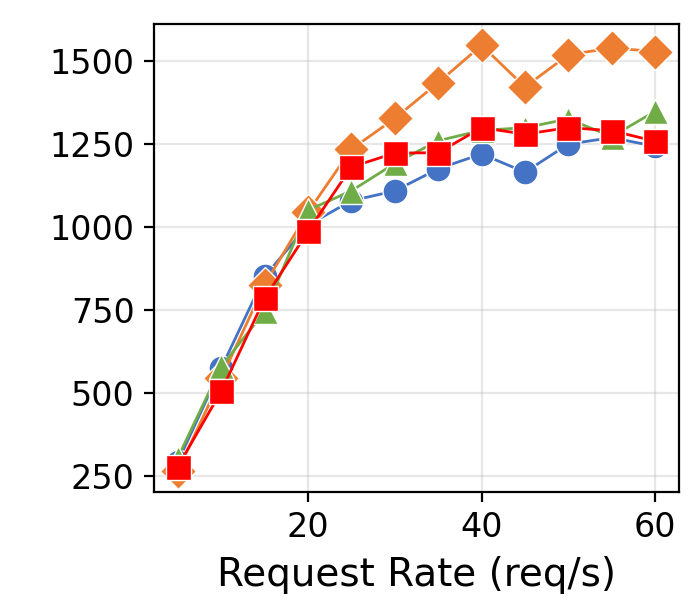}

\vspace{2pt}
\noindent\makebox[0.249\textwidth][c]{\small (a) ShareGPT, unfused}%
\makebox[0.249\textwidth][c]{\small (b) Alpaca, unfused}%
\makebox[0.249\textwidth][c]{\small (c) ShareGPT, fused}%
\makebox[0.249\textwidth][c]{\small (d) Alpaca, fused}
\caption{Throughput versus request rate under different brownout configurations: (a) ShareGPT and (b) Alpaca with unfused MoE execution; (c) ShareGPT and (d) Alpaca with fused MoE kernels}
\label{fig:exp_perf}
\end{figure*}

We further evaluate inference throughput under different brownout configurations, with results shown in Figure~\ref{fig:exp_perf}.

As shown in Figure~\ref{fig:exp_perf}, more aggressive brownout improves throughput by reducing active experts per request. Under unfused MoE, executing fewer expert instances directly lowers dispatch and communication work, so the curves separate as the grouping factor increases. The 8-way configuration reaches 475 tokens/s on ShareGPT and 460 tokens/s on Alpaca, improving over the baseline by 2.1$\times$ and 2.24$\times$ under unfused MoE execution, respectively. The difference between the two traces also illustrates the effect of workload shape: ShareGPT and Alpaca induce different prompt and generation-length distributions, so the same expert reduction does not produce identical token rates.

The fused-MoE panels provide an important control experiment. With fused kernels, all configurations converge to similar peak throughput ($\sim$1750--1800 tokens/s), because kernel fusion absorbs much of the dispatch overhead that BrownoutMoE removes in the unfused path. This does not make grouping unnecessary: the fused setting still benefits from lower memory pressure and can be combined with the same SLO controller, but its marginal throughput gain is smaller. Taken together, the four panels show that BrownoutMoE is most valuable when expert dispatch and communication, rather than fused arithmetic, dominate the serving critical path.

\subsection{Scalability and Latency}\label{sec:scalability}

\begin{figure*}[t]
\centering
\begin{tikzpicture}[x=0.48cm, y=0.48cm]
\definecolor{c2gpu}{HTML}{ED7D31}
\definecolor{c4gpu}{HTML}{70AD47}
\definecolor{c8gpu}{HTML}{FF0000}
\draw[color=c2gpu, line width=0.8pt] (0,0) -- (1.2,0);
\fill[c2gpu] (0.6,0) ++(0pt,2.2pt) -- ++(2.2pt,-2.2pt) -- ++(-2.2pt,-2.2pt) -- ++(-2.2pt,2.2pt) -- cycle;
\node[right, font=\scriptsize] at (1.4,0) {2-GPU};
\draw[color=c4gpu, line width=0.8pt] (5.5,0) -- (6.7,0);
\fill[c4gpu] (6.1,0) -- ++(2pt,2.8pt) -- ++(2pt,-2.8pt) -- ++(-4pt,0pt) -- cycle;
\node[right, font=\scriptsize] at (6.9,0) {4-GPU};
\draw[color=c8gpu, line width=0.8pt] (11.0,0) -- (12.2,0);
\fill[c8gpu] (11.6,0) ++(-1.8pt,-1.8pt) rectangle ++(3.6pt,3.6pt);
\node[right, font=\scriptsize] at (12.4,0) {8-GPU};
\end{tikzpicture}

\vspace{2pt}
\includegraphics[width=0.247\textwidth]{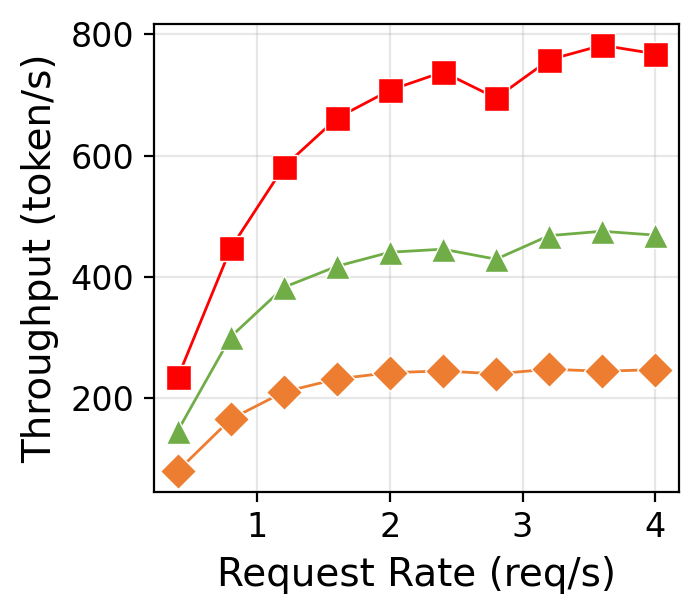}%
\hspace{0.002\textwidth}%
\includegraphics[width=0.247\textwidth]{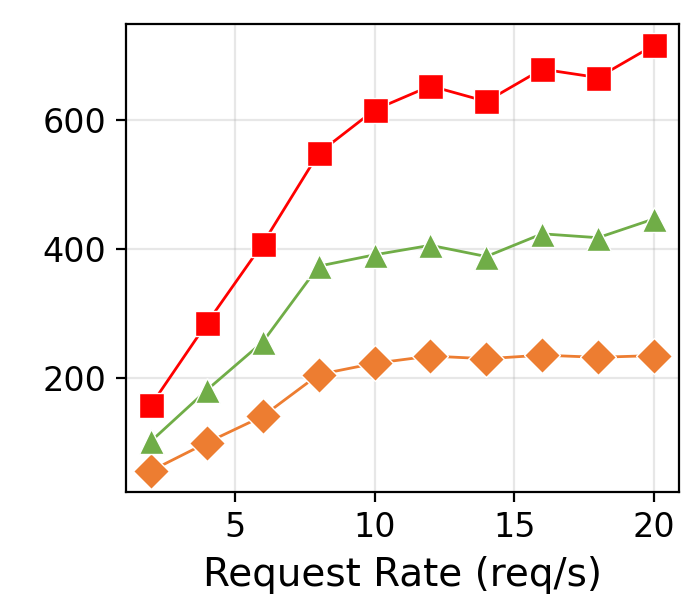}%
\hspace{0.002\textwidth}%
\includegraphics[width=0.247\textwidth]{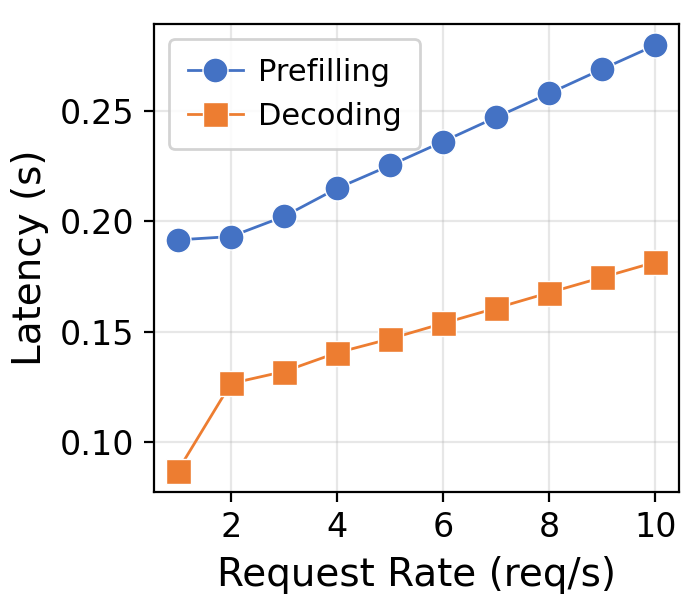}%
\hspace{0.002\textwidth}%
\includegraphics[width=0.247\textwidth]{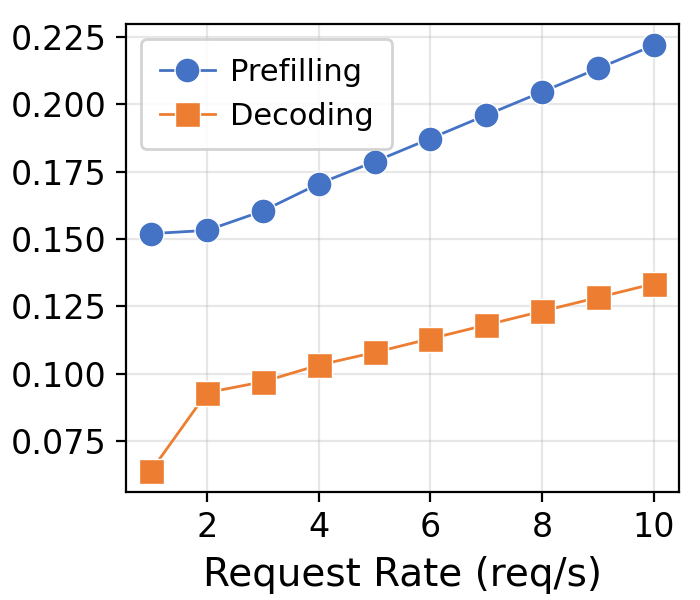}

\vspace{2pt}
\noindent\makebox[0.249\textwidth][c]{\small (a) ShareGPT}%
\makebox[0.249\textwidth][c]{\small (b) Alpaca}%
\makebox[0.249\textwidth][c]{\small (c) ShareGPT}%
\makebox[0.249\textwidth][c]{\small (d) Alpaca}
\caption{System scalability and latency on ShareGPT and Alpaca: (a), (b) throughput versus request rate with 2, 4, and 8 GPUs under tensor parallelism; (c), (d) prefill and decode latency versus request rate with 8 GPUs. The legend above the panels applies to (a) and (b)}
\label{fig:exp_scalability}
\end{figure*}

Figure~\ref{fig:exp_scalability} evaluates scalability under different GPU counts with tensor parallelism in unfused MoE mode.

As shown in Figure~\ref{fig:exp_scalability}(a) and (b), throughput increases with request rate until the available GPU and communication capacity is saturated. On ShareGPT, 4-GPU reaches about 1.9$\times$ the 2-GPU peak, and 8-GPU further delivers 1.65$\times$ over 4-GPU. The sublinear scaling is expected for a distributed MoE service: additional GPUs provide more expert capacity, but token exchange and synchronization increasingly contribute to the critical path. BrownoutMoE reduces the amount of expert work per request before this communication stage, which explains why its gains are most visible at the higher-load portions of the curves.

Figure~\ref{fig:exp_scalability}(c) and (d) show gradual latency growth under 8 GPUs rather than an abrupt instability. At low request rates, the system is compute-bound and the curves remain close; as the arrival rate increases, queueing and all-to-all contention become more important. The smooth increase indicates that the SLO-aware path has room to reduce the threshold before the service crosses its target. This is the operational role of the controller: it converts the static efficiency-quality trade-off into a workload-dependent policy instead of committing every request to zero- or full-brownout execution.

The scalability and latency results should be read together with the accuracy rows. Increasing the GPU count does not change the grouping artifact or its distillation quality; it changes only the amount of serving capacity available to execute that artifact. Thus, the offline structure decision and the online resource decision remain separable, while the threshold provides a controlled response when the workload approaches saturation.

\subsection{Per-Second Latency under Adaptive Control}\label{sec:timely}

\begin{figure*}[t]
\centering
\includegraphics[width=0.48\textwidth]{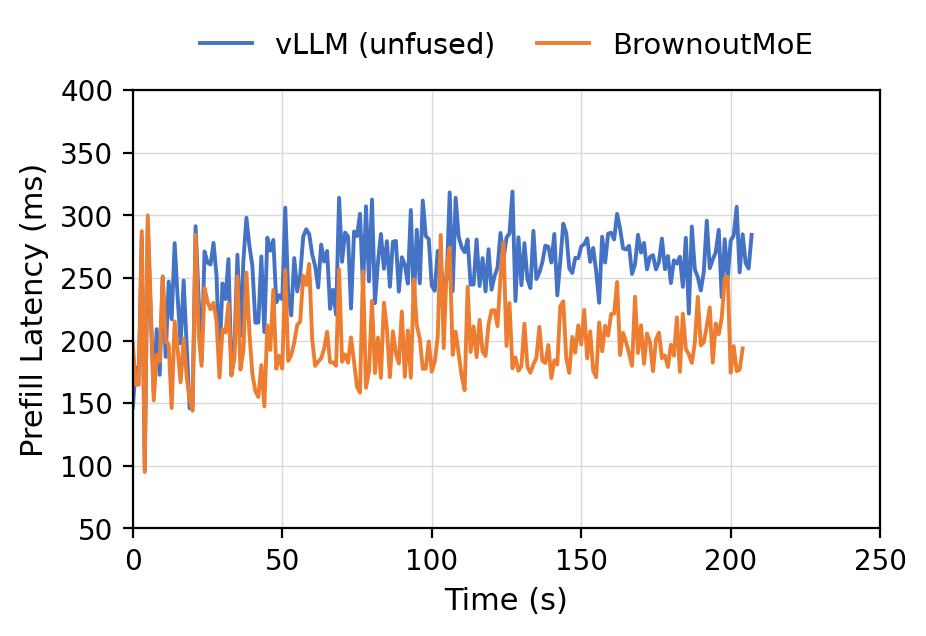}%
\hspace{0.02\textwidth}%
\includegraphics[width=0.48\textwidth]{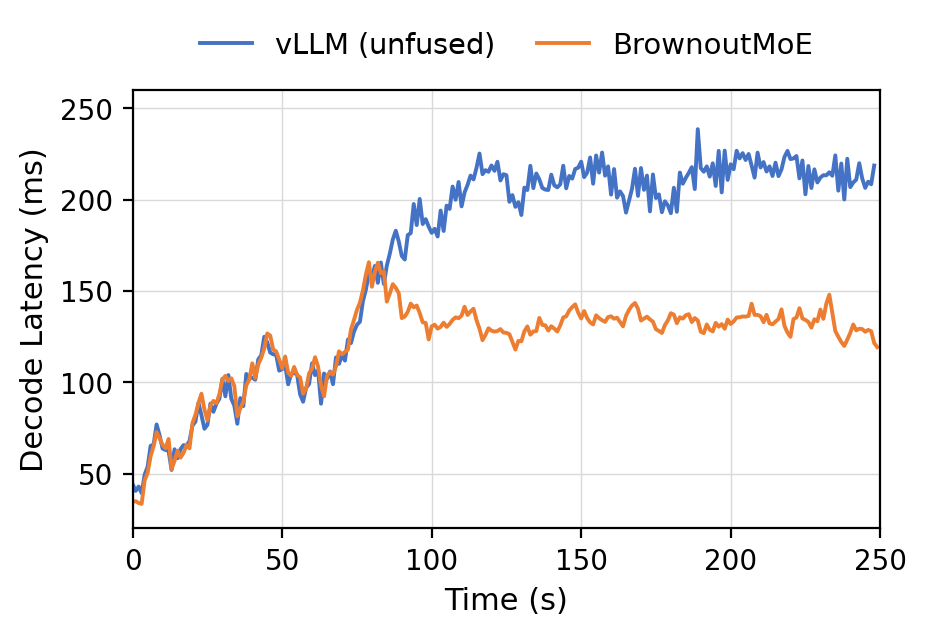}

\vspace{2pt}
\noindent\makebox[0.48\textwidth][c]{\small (a) Prefill latency}%
\hspace{0.02\textwidth}%
\makebox[0.48\textwidth][c]{\small (b) Decode latency}
\caption{Per-second average (a) prefill and (b) decode latency on ShareGPT under continuously increasing load: the unfused vLLM baseline versus BrownoutMoE with adaptive brownout control. The first warm-up second of the prefill measurements is excluded}
\label{fig:timely_latency}
\end{figure*}

To observe the controller behavior at fine granularity, we further record the average prefill and decode latency of every elapsed second during a ShareGPT run in which the offered load increases continuously. Figure~\ref{fig:timely_latency} compares the unfused vLLM baseline with BrownoutMoE under the SLO-aware controller.

The two decode curves track each other closely during the first minute, when the service is still underloaded and the controller keeps the threshold at 1.0, which confirms that the controller does not sacrifice quality while latency pressure is mild. As the arrival rate rises, the baseline climbs from roughly 80 ms to a 200--225 ms plateau and peaks at about 239 ms, whereas the controller lowers the brownout threshold from 1.0 to approximately 0.5 and holds BrownoutMoE between 125 and 145 ms. Over the saturated final 150 seconds, the mean decode latency drops from about 216 ms to 133 ms (a 38\% reduction), and the peak is reduced by about 30\%. The prefill panel shows the same pattern: the baseline oscillates around 257 ms with excursions up to about 319 ms, while BrownoutMoE remains near 200 ms, a 22\% lower sustained prefill latency. These traces complement the request-rate sweep in Figure~\ref{fig:exp_scalability}: rather than a single operating point, they show the complete temporal path of the service, and the smooth transition between the tracking and the controlled regime indicates that the threshold adaptation is stable enough for production serving.

\subsection{Generalizability}\label{sec:general}

BrownoutMoE uses routing statistics and post-distillation error rather than model-specific modifications, making the grouping pipeline applicable to other sparse MoE models. The required interfaces are a router that exposes expert assignments, an expert-output calibration pass, and a serving runtime that can select between original and united experts. Models with different expert counts or top-$K$ values change the capacity constraint and calibration cost, but not the objective, the GRPO search structure, or the SLO control loop.

\section{Discussion and Implications}\label{sec:discussion}

This section interprets the evaluation results and discusses their deployment implications, rather than introducing another conclusion. The experiments show that expert grouping is a model-structure decision that should be optimized together with the quality objective of the united expert. The throughput plots show that the runtime benefit comes from executing fewer expert instances, while the three accuracy figures show that the benefit is not free when groups are formed arbitrarily. BrownoutMoE improves this trade-off by spending computation offline to select groups that are easier to distill. Because the online operator and the original router are unchanged, the gain can be combined with independent kernel, batching, and prefill/decode optimizations.

\textbf{Why structure-aware grouping works.} A united expert is a single student that must reproduce a mixture of member functions. When the grouped experts are behaviorally similar, their outputs occupy a compact region of the function space, so a short-horizon student can approximate the whole group with low routing-weighted error. When dissimilar experts share a student, one parameter set must cover incompatible activation regions, and the approximation error compounds across layers. Because the GRPO reward is the post-distillation error itself, the search directly optimizes the quantity that determines deployed accuracy, which explains why the learned maps outperform index-order grouping at every grouping factor.

\textbf{When BrownoutMoE is beneficial.} The margin over sequential grouping grows with the grouping factor: at 2-way, many expert pairs are already compatible, so both grouping strategies distill well; at 8-way, index proximity becomes a poor proxy for functional compatibility, and the learned grouping retains a clear advantage. The benefit is also larger for structure-sensitive datasets such as C-Eval and OBQA, whose tasks depend on specialized experts, and for serving paths where expert dispatch and communication dominate the critical path, as in the unfused MoE execution studied here.

\textbf{When it is less beneficial.} On datasets whose experts are largely interchangeable, such as PIQA and COPA at small grouping factors, both grouping strategies already approach the Zero Brownout reference, so the additional gain from learning the grouping is marginal. With fused MoE kernels, dispatch overhead is largely absorbed, so all configurations converge in peak throughput and the structural benefit shrinks to memory-pressure relief. Finally, when latency pressure is mild and the brownout threshold stays high, most requests remain on the original experts and the grouping artifact is rarely exercised; the online controller, rather than the offline grouping, is then the active lever.

There are three practical implications. First, grouping should be learned per layer rather than copied across layers: the routing distribution and specialization pattern vary substantially with depth, and the routing-weighted objective naturally gives high-traffic layers more influence during model preparation. Second, the grouping artifact should be versioned with the model checkpoint and calibration profile. A workload with a different domain or prompt distribution can change expert frequencies, so a deployment may regenerate the artifact periodically while retaining the same serving interface. The SLO Analyzer can continue to adjust the online threshold between artifact refreshes. Third, for operators of shared GPU clusters, the separation between offline artifacts and online control turns grouping into a model-preparation task: artifacts can be generated on spare capacity and rolled out or rolled back per model version without interrupting the serving path.

The current study also has limitations. The evaluation uses one Qwen1.5-MoE model, deployments with up to eight GPUs, and a fixed set of downstream benchmarks and LLM service traces. Short-horizon distillation is an effective search proxy, but it can miss rare activation regions that appear only in production. In addition, the table compares methods at the system level; it does not claim that BrownoutMoE replaces kernel or routing innovations. Future work should evaluate larger and multilingual MoE models on multi-node GPU clusters with expert parallelism, add drift-aware calibration, and study joint optimization of grouping, expert placement, and request-level admission control.

\section{Conclusion and Future Work}\label{sec:conclusion}

We presented BrownoutMoE, a structure-aware expert grouping framework for efficient and accurate MoE inference services. The central observation is that sparse activation alone does not remove the structural cost of expert fragmentation: a router may activate only a few experts for each token, while the serving system still pays for dispatch, communication, and independently executed expert functions. BrownoutMoE addresses this problem at the model-structure level by learning which experts can be represented by a shared united expert and by separating offline structure optimization from online brownout control.

The proposed pipeline combines routing-aware calibration, similarity-based initialization, GRPO exploration, and grouping-consistent knowledge distillation. The reward is computed from the post-distillation error rather than from an index-based or parameter-only proxy, so the learned grouping is directly tied to the quality of the deployable united experts. At runtime, the original router and batching interface remain unchanged. The SLO Analyzer adjusts only the brownout threshold, allowing the service to use more united-expert computation during overload and to recover original-expert computation when latency pressure subsides.

Across Qwen1.5-MoE-A2.7B workloads, BrownoutMoE reduces accuracy degradation by up to 71.4\% relative to sequential grouping, and its brownout path improves unfused throughput by up to 2.24$\times$ over a vLLM baseline. The improvement is most pronounced under aggressive 4-way and 8-way grouping, where arbitrary expert combinations create the largest distillation error. These results show that grouping quality is a first-class systems variable: kernel fusion, scheduling, and routing optimizations remain valuable, but their benefits can be limited when the model artifact itself contains incompatible united experts.

The current evaluation focuses on one MoE checkpoint, a fixed calibration distribution, and a limited set of downstream benchmarks and LLM service traces. Future work should study larger, multilingual, and continuously updated MoE models deployed across multi-node GPU clusters; refresh grouping artifacts under distribution drift; and jointly optimize expert placement, admission control, and brownout thresholds. More broadly, BrownoutMoE provides a reusable interface for turning offline knowledge of expert behavior into an online quality--latency control policy for dynamic LLM services running on parallel and distributed GPU infrastructure.

\backmatter

\bibliography{BrownoutMoE}

\section*{Statements and Declarations}

\bmhead{Funding}
This work was supported by the National Key Research and Development Program of China (Grant No. 2026YFE0199800), the National Natural Science Foundation of China (Grant No. 62572462), the Guangdong Science and Technology Cooperation Project (Grant No. 2025A0505020065), the Guangdong Basic and Applied Basic Research Foundation (Grant No. 2024A1515010251), the Key Research and Development and Technology Transfer Program of Inner Mongolia Autonomous Region (Grant No. 2025YFHH0110), and the Shenzhen Science and Technology Program (Grant Nos. JCYJ20240813155810014 and ZDYJ20251211121533004).

\bmhead{Competing interests}
The authors have no relevant financial or non-financial interests to disclose.

\bmhead{Author contributions}
All authors participated in extensive discussions and contributed to the conceptualization, analysis, and finalization of this manuscript. Yi Ding conceived the research idea, implemented the BrownoutMoE framework, conducted the experiments, analyzed the results, and prepared the initial manuscript. Minxian Xu provided overall supervision and guidance on the research methodology, system design, and manuscript development. Zhengxin Fang assisted with implementation, experimental evaluation, data analysis, and manuscript revision. Kejiang Ye contributed critical insights into the application scenarios, serving-system design, and scalability considerations, and helped refine the manuscript. Chengzhong Xu provided senior guidance on the research direction and experimental interpretation, and contributed to the final revision of the manuscript.

\bmhead{Data availability}
The benchmark datasets and serving traces used in this study (PIQA, COPA, C-Eval, OpenBookQA, ShareGPT, and Alpaca) are publicly available from the sources cited in this article. Other data generated during the current study are available from the corresponding author on reasonable request.

\bmhead{Code availability}
The source code of BrownoutMoE is available at \url{https://github.com/done-77/BrownoutMoE}.

\bmhead{Ethics approval and consent to participate}
Not applicable.

\bmhead{Consent for publication}
Not applicable.

\end{document}